\documentclass[prd,nofootinbib,preprint,showpacs,superscriptaddress]{revtex4}

\usepackage{amsmath}

\newcommand{\axi}[1]{|\xi_{#1}|}
\newcommand{\hxi}[1]{\hat{\xi}_{#1}}

\newcommand{\Br}[1]{\mathcal{B}(#1)}
\newcommand{\su}{SU(3)}
\newcommand{\Bd}{B^0_d}
\newcommand{\Bs}{B^0_s}
\newcommand{\Bu}{B^+_u}
\newcommand{\Kb}{\overline{K^0}}

\newcommand{\Ksb}{\overline{K^{*0}}}

\renewcommand{\Re}{\mathcal{R}e}
\renewcommand{\Im}{\mathcal{I}m}

\begin{document}

\preprint{{\vbox{\hbox{}\hbox{}\hbox{}
\hbox{hep-ph/0509125}}}}

\vspace*{1.5cm}

\title{Using $SU(3)$ Flavor to Constrain the CP Asymmetries in $B \to
  PP,VP,VV$ Decays involving $b \to s$ Transitions}

\author{Guy Raz}\email{guy.raz@weizmann.ac.il}
\affiliation{Department of Particle Physics,
  Weizmann Institute of Science, Rehovot 76100, Israel}


\vspace{2cm}
\begin{abstract}
  We use the approximate $\su$ flavor symmetry of the strong
  interaction to put bounds on the CP asymmetries in $b \to s$
  decays of the $\Bd$ and $\Bu$ mesons. We extend the work
  of~\cite{Grossman:2003qp} to include all relevant $B \to PP$, $B \to
  VP$ and $B \to VV$ decays. We obtain the strongest
  constraints from current data, and provide a list of
  $\su$ relations which can be used when future data is obtained.
\end{abstract}

\maketitle

\section{Introduction}
\label{sec:introduction}

This work concerns the CP asymmetries in two-body $b \to s$ decays of
$B$ mesons:
\begin{equation}
  \label{eq:1}
  \mathcal{A}_f(t) \equiv \frac{\Gamma(\overline{B}^0(t) \to f) -
    \Gamma(B^0(t) \to f)}{\Gamma(\overline{B}^0(t) \to f) +
    \Gamma(B^0(t) \to f)} \;.
\end{equation}
For a final state $f$ which is a CP eigenstate, the CP asymmetry is
time dependent,
\begin{equation}
  \label{eq:2}
  \mathcal{A}_f(t) = -C_f\,\cos(\Delta m_B t) + S_f\, \sin(\Delta m_B t)\;,
\end{equation}
while for a flavor specific final state the CP asymmetry is time
independent,
\begin{equation}
  \label{eq:3}
  \mathcal{A}_f(t) = \mathcal{A}_f\;.
\end{equation}

In the standard model, the $b \to s$ transition amplitudes are naively
dominated by a single weak phase. A contribution from a second weak
phase is CKM suppressed by order $\mathcal{O}(\lambda^2)$.
Consequently, the corresponding CP asymmetries are naively expected to
fulfill $-\eta_f\,S_f \approx S_{\psi K_S}$ and $C_f,\,\mathcal{A}_f
\approx 0$ up to order of a few percent. A deviation from those
expectations can serve as a signal for new physics while an agreement
with them would imply yet another success of the standard model.

The word ``naively'' is used here for a reason. While the CKM
suppression factors are well known, the amplitude also depends on
hadronic matrix elements for which there is no fundamental theory
which is proven to a high level of precision. In order to estimate the
allowed deviation of the CP asymmetries from the naive expectation
within the SM, one needs to calculate these hadronic matrix elements
(or at least, the ratios between them).

In this work we follow the method of~\cite{Grossman:2003qp} and use
the approximate $\su$ flavor symmetry of the strong interaction to
bound the ratio between the relevant terms in the $b \to s$ decay
amplitudes. We extend the analysis of~\cite{Grossman:2003qp} to
include $b \to s$ decays in all two-body $B \to PP$, $B \to VP$ and $B
\to VV$ decays. Here, $B$ stands for either $\Bd$ or $\Bu$, $P$ stands
for the pseudo-scalar meson nonet and $V$ stands for the vector meson
nonet. The inclusion of $B \to VV$ modes requires some adjustment of
the arguments in~\cite{Grossman:2003qp} so that they will apply to the
case of non-single final state (similar in spirit
to~\cite{Engelhard:2005hu,Engelhard:2005ky}). The currently available
experimental values of the CP asymmetries for these modes, taken
from~\cite{HFAG}, are collected into table~\ref{tab:BtoPP} ($B \to
PP$), table~\ref{tab:BtoVP} ($B \to VP$) and table~\ref{tab:BtoVV} ($B
\to VV$).

\begin{table}
  \centering
  \begin{tabular}{|c|c|c|c|}
    \hline
    Mode & $-\eta_f S_f$ & $C_f,\, -\mathcal{A}_f$ \\
    \hline
    \hline
    $\Bd \to \eta K_S$ & --- & --- \\
    $\Bd \to \eta' K_S$ & $0.50 \pm 0.09$ & $-0.07 \pm 0.07$ \\
    $\Bd \to \pi^0 K_S$ & $0.31 \pm 0.26$ & $-0.02 \pm 0.13$ \\
    $\Bd \to \pi^- K^+$ & n/a & $0.115 \pm 0.018$ \\
    \hline
    $\Bu \to \eta K^+$ & n/a & $0.33 \pm 0.12$ \\
    $\Bu \to \eta' K^+$ & n/a & $0.031 \pm 0.021$ \\
    $\Bu \to \pi^+ K^0$ & n/a & $0.02 \pm 0.04$ \\
    $\Bu \to \pi^0 K^+$ & n/a & $-0.04 \pm 0.04$ \\
    \hline
  \end{tabular}
  \caption{Measured CP asymmetries in $B \to PP$, $b \to s$ decays.}
  \label{tab:BtoPP}
\end{table}

\begin{table}
  \centering
  \begin{tabular}{|c|c|c|c|}
    \hline
    Mode & $-\eta_f S_f$ & $C_f,\, -\mathcal{A}_f$ \\
    \hline
    \hline
    $\Bd \to \phi K_S$ & $0.47 \pm 0.19$ & $-0.09 \pm 0.14$ \\
    $\Bd \to \omega K_S$ & $0.63 \pm 0.30$ & $-0.44 \pm 0.23$ \\
    $\Bd \to \rho^0 K_S$ & --- & --- \\
    $\Bd \to \rho^- K^+$ & n/a & $-0.17^{+0.16}_{-0.15}$ \\
    $\Bd \to K^{*0} \eta$ & n/a & $0.01 \pm 0.08$ \\
    $\Bd \to K^{*0} \eta'$  & n/a & --- \\
    $\Bd \to K^{*0} \pi^0$  & n/a & $0.01^{+0.26}_{-0.27}$ \\
    $\Bd \to K^{*+} \pi^-$  & n/a & $0.05 \pm 0.14$ \\
    \hline
    $\Bu \to \phi K^+$ & n/a & $-0.037 \pm 0.050$ \\
    $\Bu \to \omega K^+$ & n/a & $-0.02 \pm 0.07$ \\
    $\Bu \to \rho^+ K^0$ & n/a & --- \\
    $\Bu \to \rho^0 K^+$ & n/a & $-0.31^{+0.11}_{-0.12}$ \\
    $\Bu \to K^{*+} \eta$ & n/a & $-0.03^{+0.10}_{-0.11}$ \\
    $\Bu \to K^{*+} \eta'$ & n/a & --- \\
    $\Bu \to K^{*0} \pi^+$ & n/a & $0.093 \pm 0.060$ \\
    $\Bu \to K^{*+} \pi^0$ & n/a & $-0.04 \pm 0.29$ \\
    \hline
  \end{tabular}
  \caption{Measured CP asymmetries in $B \to VP$, $b \to s$ decays.}
  \label{tab:BtoVP}
\end{table}

\begin{table}
  \centering
  \begin{tabular}{|c|c|c|c|}
    \hline
    Mode & $-\eta_f S_f$ & $C_f,\, -\mathcal{A}_f$ \\
    \hline
    \hline
    $\Bd \to \phi K^{*0}$ & n/a & $0.00 \pm 0.07$ \\
    $\Bd \to \omega K^{*0}$ & n/a & --- \\
    $\Bd \to \rho^0 K^{*0}$ & n/a & --- \\
    $\Bd \to \rho^- K^{*+}$ & n/a & --- \\
    \hline
    $\Bu \to \phi K^{*+}$ & n/a & $-0.05 \pm 0.11$ \\
    $\Bu \to \omega K^{*+}$ & n/a & --- \\
    $\Bu \to \rho^+ K^{*0}$ & n/a & $0.14 \pm 0.43$ \\
    $\Bu \to \rho^0 K^{*+}$ & n/a & $-0.20^{+0.29}_{-0.32}$ \\
    \hline
  \end{tabular}
  \caption{Measured CP asymmetries in $B \to VV$, $b \to s$ decays.}
  \label{tab:BtoVV}
\end{table}

We do not consider here $b \to s$ decays of the $\Bs$ meson although
the same method applies here as well. We do include, however, $\Bs$
decay modes in the theoretical decomposition to $\su$ invariant
amplitude. We also comment on several one-to-one amplitude relations
between specific $\Bd$ and $\Bs$ decays usually due to the U-spin
subgroup of $\su$.

Using $\su$ has two weaknesses. First, this symmetry is broken by
effects of order $m_s/\Lambda_{\chi_\text{SB}} \sim 0.3$. The bounds
we obtain can therefore be violated to this order. Second, our method
often provides only a (conservative) upper bound on the deviation and
not an estimate. The actual deviation can be substantially lower than
our bounds. The advantage of our method, on the other hand, is its
hadronic model independence. Thus, methods that use the approximate
symmetries of the strong interaction~\cite{Grossman:2003qp,
  Engelhard:2005hu, Engelhard:2005ky, Chiang:2003pm, Gronau:2003kx,
  Gronau:2004hp} are complementary to methods employing direct
calculation of hadronic matrix elements within factorization related
schemes~\cite{Ali:1998eb, Beneke:2001ev, Beneke:2003zv,
  Buchalla:2005us, Beneke:2005pu}.

The structure of this work is as follows: In
section~\ref{sec:formalism} we review the formalism relevant to the
use of $\su$ in constraining the CP asymmetries in $b \to s$
transitions and comment on the strategy of this
work. Section~\ref{sec:results} is divided into three
subsections dedicated to $B \to PP$ (subsection~\ref{sec:b-to-pp}), $B
\to VP$ (subsection~\ref{sec:b-to-vp}) and $B \to VV$
(subsection~\ref{sec:b-to-vv-2}) decays. For each decay mode we give
the $\su$ relation which leads to the strongest constraint currently
available. We also provide a list of additional relations for each
mode and argue that, in the future, one of those relations is likely
to provide the strongest bound. We conclude in
section~\ref{sec:summary}. For readers who are only interested in
the current strongest constraints on the CP asymmetries, we summarized
our results in this section. In addition, we give the details on how to
derive the $\su$ reduced matrix elements relations in
appendix~\ref{sec:su3-formalism} and provide the resulting
decomposition of all physical modes in three tables. We also quote the
current available experimental branching ratios for all relevant modes
in appendix~\ref{sec:experimental-data}.

\section{Formalism}
\label{sec:formalism}

In this section we provide a short review of the formalism and
introduce the notation relevant to the use of $\su$ symmetry in $b \to
s$ decays~\cite{Grossman:2003qp,Engelhard:2005hu}. We also provide
further comments on several relevant issues.

We consider a $B \to f$ decay process involving a $b\to s$
transition, and write the decay amplitude as
\begin{equation}
  \label{eq:4}
  A_f = \lambda^s_c\,a^c_f + \lambda^s_u\,a^u_f\;.
\end{equation}
We use the compact notation $\lambda^{q'}_q \equiv V^*_{qb}V_{qq'}$.

In the context of this work, writing the amplitude as in
eq.~(\ref{eq:4}) has two important merits: First, the parameters we
use have definite behaviour under CP conjugation. In the conjugate
decay process $\overline{B} \to \overline{f}$, the CKM factors get
complex conjugated while the $a^q_f$ terms remain unchanged. The CP
asymmetries can be therefore related in a simple way to those
parameters. Second, in this way of writing we isolate the approximate
$\su$ invariant part of the amplitude, the strong factors $a^q_f$,
from the explicitly $\su$ breaking ones, the weak CKM factors.

The difficulty in interpreting the CP asymmetries lies in the fact
that there is no fundamental theory for calculating the $a^q_f$
parameters. Those are essentially hadronic matrix elements which
involve non-perturbative calculations of the strong interaction. The
effect of these hadronic parameters on CP asymmetries, however, can be
encoded, for each decay process, into a single
object~\cite{Grossman:2003qp}
\begin{equation}
  \label{eq:5}
  \xi_f \equiv
  \frac{\left|\lambda^s_u\right|}{\left|\lambda^s_c\right|}
  \frac{a^u_f}{a^c_f}\;.
\end{equation}
For a final state $f$ which is a CP eigenstate, the observed CP
asymmetries are given, to first order in $\xi_f$, by
\begin{align}
  -\eta_f\,S_f - \sin 2\beta & = 2 \cos 2\beta\, \sin \gamma\, \Re\,
  \xi_f\;, \label{eq:6} \\
  C_f &= -2\sin\gamma\, \Im\, \xi_f\;. \label{eq:7}
\end{align}
Combining the two relations~(\ref{eq:6}) and~(\ref{eq:7}) we have
\begin{equation}
  \label{eq:8}
  \left[(-\eta_f\,S_f - \sin 2\beta)/\cos 2\beta\right]^2+
  C_f^2 = 4\sin^2\gamma\,|\xi_f|^2\;.
\end{equation}
For a final state $f$ which is flavor specific, the CP asymmetry
$-\mathcal{A}_f$ is given by the same formula~(\ref{eq:7}).

$\su$ relations provide a way to constrain $\axi{f}$ and
therefore constrain the CP asymmetries. Owing to the approximate $\su$
of the strong interaction, the $a^q_f$ parameters can be expressed
using corresponding parameters of $b \to d$ decay processes,
\begin{equation}
  \label{eq:9}
  a^q_f = \sum\limits_{f'} X_{f'}\,b^q_{f'}\;,
\end{equation}
where the $b^q_{f'}$ enters the $B \to f'$ decay amplitude
\begin{equation}
  \label{eq:10}
  A_{f'} = \lambda^d_c\,b^c_{f'} + \lambda^d_u\,b^u_{f'}\;.
\end{equation}
Typically, for any mode $f$ there are many possible relations of the
form~\eqref{eq:9}. Finding the various coefficients $X_{f'}$ is the
main topic of this work.

Once a relation of the form~(\ref{eq:9}) is established, we can use
the $X_{f'}$'s to define
\begin{equation}
  \label{eq:11}
  \hxi{f} \equiv \left|\frac{V_{us}}{V_{ud}}\right|
  \left|\frac{\sum\limits_{f'} X_{f'}\,A_{f'}}{A_f}\right| \leq 
  \left|\frac{V_{us}}{V_{ud}}\right|  
  \frac{\sum\limits_{f'} \left|X_{f'}\right| \sqrt{\mathcal{B}\left( B
        \to f'\right)}}{\sqrt{\mathcal{B}\left( B \to f\right)}}\;.
\end{equation}
If experimental data yield an upper bound on $\hxi{f}$ in the range
between $\lambda^2$ and $1$ ($\hxi{f}$ is positive by definition), a
constraint on $\axi{f}$ is implied:
\begin{equation}
  \label{eq:12}
  \left|\xi_f\right| \leq \frac{\hxi{f}+\lambda^2}{1-\hxi{f}}\;.
\end{equation}
In fact, if we use the charge averaged branching ratios
in~(\ref{eq:11}), (and correspondingly replace the $|\sum_{f'}
X_{f'}\,A_{f'}|$ with $\sqrt{|\sum_{f'} X_{f'}\,A_{f'}|^2+|\sum_{f'}
  X_{f'}\,\bar{A}_{f'}|^2}$ and the $|A_f|$ with
$\sqrt{|A_f|^2+|\bar{A}_f|^2}$ in the definition for $\hxi{}$) the
constraint on $|\xi_f|$ can be slightly stronger than~(\ref{eq:12}),
depending on the value of the weak phase
$\gamma$~\cite{Engelhard:2005hu}. Keeping a conservative path, we take
the worst case scenario and make no assumptions regarding $\gamma$.

Current experiments give no data on the relative phase of the various
decay amplitudes. Since we wish to obtain an upper bound on $\xi_f$,
we must make use of the triangle inequality and add the terms
in~(\ref{eq:11}) with an absolute value. While this method indeed
guarantees that we obtain an upper bound (assuming $\su$), it also has
the potential of weakening this bound considerably.

Using this formalism, therefore, the best prospects for obtaining a
strong constraint on the CP asymmetries lies in those amplitude
relations which involve the smallest number of modes. In this work we
follow this logic and present relations with up to three modes
involved. Currently those relations do give the strongest available
constraints.

We stress that when we calculate $\hxi{f}$ from~(\ref{eq:11}) we use
either the experimental bound on the branching ratio $\Br{B \to f'}$,
if only a bound exists, or the central value, if an actual measurement
exists. The fact that we use central values can, potentially, somewhat
strengthen $\hxi{f}$. Being conservative in other respects we do not
consider this to be a significant issue.

In $B \to VV$ decays, the measured CP asymmetry and branching ratios
represent a sum over three possible final states with distinct orbital
angular momentum configuration, namely $l=0,\,1$ or $2$. Since all $B
\to VV$ decays with $b \to s$ transitions are flavor specific, the
only relevant CP asymmetry is $\mathcal{A}_f$. As was explained in
detail in~\cite{Engelhard:2005hu,Engelhard:2005ky}, an extension of
the same method can be used to constrain the CP asymmetry obtained in
the case of a sum over several final states.

Since here there are only three final states we write the expression
explicitly. We consider the three amplitudes
\begin{equation}
  \label{eq:13}
  A_{f;\,l}= \lambda^s_c\,a^c_{f;\,l} +
  \lambda^s_u\, a^u_{f;\,l}\;,
\end{equation}
with $l=0,\,1$ or $2$. The modified parameter 
\begin{equation}
  \label{eq:14}
  \xi_{f} \equiv
  \frac{\left|\lambda^s_u\right|}{\left|\lambda^s_c\right|} \frac{
    a^{c*}_{f;\,0}a^{u}_{f;\,0} + a^{c*}_{f;\,1}a^{u}_{f;\,1} +
    a^{c*}_{f;\,2}a^{u}_{f;\,2}} { \left|a^c_{f;\,0}\right|^2 +
    \left|a^c_{f;\,1}\right|^2 + \left|a^c_{f;\,2}\right|^2}\;,
\end{equation}
enters the CP asymmetry in the same way as before:
\begin{equation}
  \label{eq:15}
  \mathcal{A}_{f} = 2\,\sin \gamma\, \Im\, \xi_f\;.
\end{equation}
As can be seen, when there is only a single final state,~(\ref{eq:14})
reduces to~(\ref{eq:5}).


Taking $\mathcal{B}\left(B \to f^{(')}\right)$ in~(\ref{eq:11}) to
indicate a sum over the branching ratios of the three final states, we
find that $\axi{f}$ is constrained by the relation given in
eq.~(\ref{eq:12}). Thus the same formalism can be applied to the
multiple final states $B \to VV$.

It is worthwhile noticing the special case in which there is a
relation between a single $b \to s$ mode and a single $b \to d$
mode. Besides being a good candidates for giving the tightest bounds,
if there is a measurement, such a relation may allow for actual
extraction of the hadronic terms and their relative phase from three
observables.

To be specific, consider that, for a CP eigenstates $f$, we have the
relation
\begin{equation}
  \label{eq:16}
  a^q_f=b^q_{f'}\;.
\end{equation}
If we measure the rate $\Br{B \to f}$, and the two CP asymmetries
$C_f$ and $S_f$, we can predict the rate $\Br{B \to f'}$ and the CP
asymmetries $C_{f'}$ and $S_{f'}$ (given that CKM factors are
known). For example, in many cases, the U-spin subgroup of $\su$ gives
such a relation between $\Bd$ and $\Bs$ decay. When accurate
experimental data will be available, this idea can be used to predict
the expected decay rate for various $\Bs$ decay mode, within the $\su$
symmetry approximation. 

Our last comment for this section concerns the issue of $\eta - \eta'$
and $\phi - \omega$ mixing. 
In this work, as was done in~\cite{Grossman:2003qp}, we do not assume
$\su$ to be an approximate symmetry in the $\eta-\eta'$ and
$\phi-\omega$ mixing. Instead, we use the phenomenological description
of the mixing and apply $\su$ relation to each component individually
(although, for $\eta$--$\eta'$ mixing the $\su$ breaking is small
enough that one could just use $\su$ straightforwardly and identify
$\eta=\eta_8$ and $\eta'=\eta_1$). In this way we are able to apply
$\su$ symmetry to decay processes without assuming $\su$ symmetry in
the masses.

While there is still room left for better theoretical understanding,
we make in this work a distinction between the physics governing
masses and mixing and the physics which governs decays. A large
breaking effect in the masses does not imply by itself a similar
breaking in decays. It is possible that, for some unknown reason,
large $\su$ breakings in the decays do occur, but currently no data
suggest this to be the case.

\section{Results}
\label{sec:results}

A good way to obtain the $\su$ relations in the form of~(\ref{eq:9})
is to write all decay processes using $\su$ invariant matrix
elements. In appendix~\ref{sec:su3-formalism} we give full details how
this is done. We also provide the result of the calculation in the
form of three tables which are relevant to any possible two-body decay
of a $B$ meson into pseudo-scalars and vectors. The tables we obtain
match those in~\cite{Grossman:2003qp} and are extended to include
$\Bs$ modes.

Using the tables it is next possible to find amplitude relations
between physical modes. In what follows we do exactly that for all
$b \to s$ process.

\subsection{$B \to PP$ modes.}
\label{sec:b-to-pp}

There are eight $b \to s$ decay modes involving final states with two
nonet pseudo-scalars. 

\subsubsection{$\Bd \to \eta K^0$ and $\Bd \to \eta' K^0$}
\label{sec:bd-to-eta-1}

These two decays were specifically studied
in~\cite{Grossman:2003qp}. We provide here an update of experimental
data as well as additional comments.

We use $\eta-\eta'$ mixing of the form
\begin{equation}
  \label{eq:17}
  \begin{aligned}
    \eta &= \eta_1\,\sin\theta_{\eta\eta'} +
    \eta_8\,\cos\theta_{\eta\eta'}\;, \\ 
    \eta' &= \eta_1\,\cos\theta_{\eta\eta'} -
    \eta_8\,\sin\theta_{\eta\eta'}\;. 
  \end{aligned}
\end{equation}
We take $\theta_{\eta\eta'}=20^{\circ}$ to be the mixing
angle~\cite{Eidelman:2004wy}.

According to our framework, in order to derive the amplitude relations
for $\Bd \to \eta^{(')} K^0$~\footnote{Note that by definition we have
  $\hxi{B \to x K^0}=\hxi{B \to x K_S}$.}, we need to discuss
separately $\Bd \to \eta_1 K^0$ and $\Bd \to \eta_8 K^0$. The $\Bd \to
\eta_1 K^0$ relations are simpler and can be obtained from
table~\ref{tab:B-to-S_1M_8}. There is a single
relation involving just one amplitude
\begin{equation}
  \label{eq:18}
  a^q_{\Bd \to \eta_1 K^0} = b^q_{\Bs \to \eta_1 \Kb}\;.
\end{equation}
There is one relation which involves two modes (neither involves
$\Bs$)
\begin{equation}
  \label{eq:19}
  a^q_{\Bd \to \eta_1 K^0} = \sqrt{\frac{3}{2}}\,b^q_{\Bd \to \eta_1
    \eta_8}- \sqrt{\frac{1}{2}}\,b^q_{\Bd \to \eta_1 \pi^0}\;.
\end{equation}
There are no relations involving three amplitudes or more.

The $\Bd \to \eta_8 K^0$ relations are obtained using
table~\ref{tab:B-to-M_8N_8}. There are two relations involving just
one amplitude. They both involve a $\Bs$ decay:
\begin{align}
  \label{eq:20}
  a^q_{\Bd \to \eta_8 K^0} & = b^q_{\Bs \to \eta_8 \Kb}\;,
  \\
  \label{eq:21}
  a^q_{\Bd \to \eta_8 K^0} & = \sqrt{\frac{1}{3}}\,b^q_{\Bs \to \pi^0 \Kb}\;.
\end{align}

Combining relations~(\ref{eq:18}) and~(\ref{eq:20}) we get the U-spin
relations which, for the physical $\eta$ and $\eta'$, imply
\begin{align}
  \label{eq:22}
  a^q_{\Bd \to \eta K^0} & = b^q_{\Bs \to \eta \Kb}\;,
  \\
  \label{eq:23}
  a^q_{\Bd \to \eta' K^0} & = b^q_{\Bs \to \eta' \Kb}\;.
\end{align}
We now demonstrate the power of such single amplitude relations. We
consider the relation~(\ref{eq:23}). Using the three measured
observables~\cite{HFAG}, $S_{\Bd \to \eta' K_S}$, $C_{\Bd \to \eta'
  K_S}$ and $\Br{\Bd \to \eta' K_S}$, together with the CKM
parameters~\cite{Charles:2004jd}, we solve for the hadronic part of
the amplitude. We use the $\su$ relation to calculate the expected
values of the observables in the $\Bs \to \eta' K_S$ decay. We get
\begin{align}
  \label{eq:24}
  S_{\Bs \to \eta' K_S} & \approx -0.46 \pm 0.29\;, \\
  \label{eq:25}
  C_{\Bs \to \eta' K_S} & \approx +0.18 \pm 0.11\;, \\
  \label{eq:26}
  2\,\Br{\Bs \to \eta' K_S} & \approx (29 \pm 18)\times 10^{-6}\;.
\end{align}
The resulting distribution is not normal. The factor of two
in~(\ref{eq:26}) is due to the $K$ mixing.

Going back to listing the relations, there are two relations involving
two amplitudes. One of them involves $\Bs$ and is of less interest
here. The other is
\begin{equation}
  \label{eq:27}
  a^q_{\Bd \to \eta_8 K^0} = \frac{1}{\sqrt{6}}\, b^q_{\Bd \to K^+ K^-} -
  \frac{1}{\sqrt{3}}\, b^q_{\Bd \to \pi^0 \pi^0}\;.
\end{equation}
Thus, the most ``economical'' relation is obtained from
combining~(\ref{eq:19}) and~(\ref{eq:27}):
\begin{align}
  \begin{split}
    a^q_{\Bd \to \eta' K^0} &= -\frac{s}{\sqrt{6}}\, b^q_{\Bd \to K^+
      K^-} + \frac{s}{\sqrt{3}}\,b^q_{\Bd \to \pi^0 \pi^0} + \sqrt{3}\,
    c^2\,s\,b^q_{\Bd \to \eta \eta} - \sqrt{3}\,
    c^2\,s\,b^q_{\Bd \to \eta' \eta'} \\
    & \qquad - \frac{c\,s}{\sqrt{2}}\,b^q_{\Bd \to \eta \pi^0} -
    \frac{c^2}{\sqrt{2}}\,b^q_{\Bd \to \eta' \pi^0}
    +\sqrt{\frac{3}{2}}\left(c^3-c\,s^2\right)\, b^q_{\Bd \to \eta
      \eta'}\;,
  \end{split} \label{eq:28} \\
  \begin{split}
    a^q_{\Bd \to \eta K^0} &= \frac{c}{\sqrt{6}}\, b^q_{\Bd \to K^+
      K^-} -\frac{c}{\sqrt{3}}\,b^q_{\Bd \to \pi^0 \pi^0} + \sqrt{3}\,
    c\,s^2\,b^q_{\Bd \to \eta \eta} - \sqrt{3}\,
    c\,s^2\,b^q_{\Bd \to \eta' \eta'} \\
    & \qquad - \frac{s^2}{\sqrt{2}}\,b^q_{\Bd \to \eta \pi^0} -
    \frac{c\,s}{\sqrt{2}}\,b^q_{\Bd \to \eta' \pi^0}
    +\sqrt{\frac{3}{2}}\left(c^2\,s-s^3\right)\, b^q_{\Bd \to \eta
      \eta'}\;,
  \end{split} \label{eq:29}
\end{align}
where we use $c\equiv \cos \theta_{\eta\eta'}$ and $s \equiv \sin
\theta_{\eta\eta'}$.

Substituting the coefficient from the relations~(\ref{eq:28})
in~(\ref{eq:11}), and using the experimental branching ratios, we get
\begin{equation}
  \hxi{\Bd \to \eta' K^0} \leq 0.17\;, \label{eq:30}
\end{equation}
which means using~(\ref{eq:12})
\begin{equation}
  \label{eq:31}
  \axi{\Bd \to \eta' K^0} \leq 0.26\;.
\end{equation}
A bound on $\hxi{\Bd \to \eta K^0}$ cannot be obtained since the
branching ratio for this mode is only bounded and not yet measured. We
therefore have no knowledge on how small the denominator
in~(\ref{eq:11}) is. We can, however, use the
relation~(\ref{eq:29}) and write the bound on $\hxi{\Bd \to
  \eta K^0}$ as a function of this branching ratio:
\begin{equation}
  \label{eq:32}
  \hxi{\Bd \to \eta K^0} \leq \frac{0.63}{\sqrt{10^6 \times \Br{\Bd
        \to \eta K^0}}}\;. 
\end{equation}
We note that already at the current upper bound, $\Br{\Bd \to \eta
  K^0}\leq 1.9\times 10^{-6}$, we get a rather weak bound on $\hxi{\Bd
  \to \eta K^0}$.

Interestingly, $\su$ relation can give some information on $\Br{\Bd
  \to \eta K^0}$. We write the following relation between $b \to s$
amplitudes
\begin{equation}
  \label{eq:33}
  a^q_{\Bd \to \eta_8 K^0} = \frac{1}{\sqrt{3}}\,a^q_{\Bd \to \pi^0 K^0}\;.
\end{equation}
Combined with $\eta-\eta'$ mixing we get the relation
\begin{equation}
  \label{eq:34}
  a^q_{\Bd \to \eta K^0} = \tan \theta_{\eta \eta'}\, a^q_{\Bd \to \eta'
    K^0} +
  \frac{1}{\sqrt{3}\,\cos \theta_{\eta \eta'}}\,a^q_{\Bd \to \pi^0
    K^0}\;.
\end{equation}
The complex phase between the two amplitudes on the RHS of
eq.~\eqref{eq:34} is not determined by experimental data, and
therefore no exact value for the $\Bd \to \eta K^0$ amplitude can be
obtained. However, we can use the relation to obtain the bounds
\begin{equation}
  \label{eq:35}
  0.66 \times 10^{-6} \lesssim \Br{\Bd \to \eta K^0} \lesssim 25
  \times 10^{-6}\;. 
\end{equation}
We therefore conclude that the branching ratio is expected to within a
factor of $3$ of the current experimental bound (see
eq.~(\ref{eq:147})).

There are nine relations involving $\Bd \to \eta_8 K^0$ with three
other amplitudes, but two of them involve $\Bs$ decays and we do not
show them. We list the remaining seven relations and the implied
constrains on $\hat{\xi}_{\Bd \to \eta^{(')} K^0}$. The constrains are
obtained by combining each relation with~(\ref{eq:19}) and rotating to
physical modes. We do not give here the explicit expression in terms
of physical states which can be easily obtained by substitution. (One
should take care, though, to properly normalize final states with two
identical mesons. See appendix~\ref{sec:su3-formalism} for details.)
\begin{multline}
  \label{eq:36}
  a^q_{\Bd \to \eta_8 K^0} = \frac{1}{\sqrt{6}}\, b^q_{\Bd \to K^+
    K^-} + \frac{1}{\sqrt{3}}\,b^q_{\Bu \to \pi^+ \pi^0} -
  \frac{1}{\sqrt{6}}\, b^q_{\Bd \to \pi^+\pi^-}\;, \\
  \Longrightarrow\;\;\;\;\;\; \hxi{\Bd \to \eta' K^0} \leq
  0.18\;,\;\;\;\;\; \hxi{\Bd \to \eta K^0} \leq \frac{0.96}{\sqrt{10^6
      \times \Br{\Bd \to \eta K^0}}}\;,  
\end{multline} 
\begin{multline}
  \label{eq:37}
  a^q_{\Bd \to \eta_8 K^0} = \sqrt{3}\, b^q_{\Bd \to \eta_8 \eta_8} -
  \sqrt{\frac{2}{3}}\,b^q_{\Bd \to K^0 \Kb} -
  \frac{1}{\sqrt{6}}\, b^q_{\Bd \to K^+ K^-}\;, \\
  \Longrightarrow\;\;\;\;\;\; 
  \hxi{\Bd \to \eta' K^0} \leq 0.17\;,\;\;\;\;\;  \hxi{\Bd \to \eta K^0} \leq
  \frac{1.01}{\sqrt{10^6 \times \Br{\Bd \to \eta K^0}}} \;,
\end{multline}
\begin{multline}
  \label{eq:38}
  a^q_{\Bd \to \eta_8 K^0} = \frac{1}{\sqrt{6}}\, b^q_{\Bd \to K^0 \Kb} -
  \frac{1}{\sqrt{3}}\,b^q_{\Bd \to \pi^0 \pi^0} +
  \frac{1}{\sqrt{2}}\, b^q_{\Bd \to \eta_8 \pi^0}\;, \\
  \Longrightarrow\;\;\;\;\;\; 
  \hxi{\Bd \to \eta' K^0} \leq 0.18\;,\;\;\;\;\;  \hxi{\Bd \to \eta K^0} \leq
  \frac{0.95}{\sqrt{10^6 \times \Br{\Bd \to \eta K^0}}} \;,
\end{multline}
\begin{multline}
  \label{eq:39}
  a^q_{\Bd \to \eta_8 K^0} = -\frac{1}{\sqrt{6}}\, b^q_{\Bd \to K^0 \Kb} -
  \frac{1}{2\sqrt{3}}\,b^q_{\Bd \to \pi^0 \pi^0} +
  \frac{\sqrt{3}}{2}\, b^q_{\Bd \to \eta_8 \eta_8}\;, \\
  \Longrightarrow\;\;\;\;\;\; 
  \hxi{\Bd \to \eta' K^0} \leq 0.17\;,\;\;\;\;\;  \hxi{\Bd \to \eta K^0} \leq
  \frac{0.65}{\sqrt{10^6 \times \Br{\Bd \to \eta K^0}}} \;,
\end{multline}
\begin{multline}
  \label{eq:40}
  a^q_{\Bd \to \eta_8 K^0} = -\sqrt{\frac{3}{2}}\, b^q_{\Bd \to K^0 \Kb}  -
  \frac{1}{\sqrt{2}}\, b^q_{\Bd \to \eta_8 \pi^0} +
  \sqrt{3}\,b^q_{\Bd \to \eta_8\eta_8}\;, \\
  \Longrightarrow\;\;\;\;\;\; 
  \hxi{\Bd \to \eta' K^0} \leq 0.17\;,\;\;\;\;\;  \hxi{\Bd \to \eta K^0} \leq
  \frac{1.20}{\sqrt{10^6 \times \Br{\Bd \to \eta K^0}}}\;,
\end{multline}
\begin{multline}
  \label{eq:41}
  a^q_{\Bd \to \eta_8 K^0} = -\sqrt{\frac{3}{2}}\, b^q_{\Bd \to K^+
    K^-}  +  {\sqrt{2}}\, b^q_{\Bd \to \eta_8 \pi^0} +
  \sqrt{3}\,b^q_{\Bd \to \eta_8\eta_8}\;, \\
  \Longrightarrow\;\;\;\;\;\; 
  \hxi{\Bd \to \eta' K^0} \leq 0.18\;,\;\;\;\;\;  \hxi{\Bd \to \eta K^0} \leq
  \frac{1.45}{\sqrt{10^6 \times \Br{\Bd \to \eta K^0}}}\;,
\end{multline}
\begin{multline}
  \label{eq:42}
  a^q_{\Bd \to \eta_8 K^0} = -\frac{\sqrt{3}}{4}\, b^q_{\Bd \to \pi^0 \pi^0}  +
  \frac{1}{2\sqrt{2}}\, b^q_{\Bd \to \eta_8 \pi^0} +
  \frac{\sqrt{3}}{4}\,b^q_{\Bd \to \eta_8\eta_8}\;, \\
  \Longrightarrow\;\;\;\;\;\; 
  \hxi{\Bd \to \eta' K^0} \leq 0.17\;,\;\;\;\;\;  \hxi{\Bd \to \eta K^0} \leq
  \frac{0.67}{\sqrt{10^6 \times \Br{\Bd \to \eta K^0}}}\;.
\end{multline}
As can be seen, there are several relations which give similar bounds
on $\hxi{}$. Typically, the value of $\hxi{}$ for these relations
differ in the next significant digits, which were rounded
here. Relation~(\ref{eq:42}) is the one which was used
in~\cite{Grossman:2003qp}. Indeed, using the available experimental
data at the time~\cite{Grossman:2003qp} was written,
relation~(\ref{eq:42}) gives a slightly stronger bound compared to the
relation~(\ref{eq:27}). We can see by comparing the bounds
in~(\ref{eq:30}) and~(\ref{eq:32}) to the ones in the list above, that
the naive expectation that ``the fewest modes the better'' is not
unreasonable.

Using this way of presenting the relations, it is easy to foresee how
future improvement in experimental data would impact the bound. In
particular, there is yet room for considerable improvement of the
bound if the branching ratios $\Br{\Bd \to \eta \eta}$, $\Br{\Bd \to
  \eta \eta'}$ and $\Br{\Bd \to \eta' \eta'}$ will have a stronger
constraint. As a demonstration, if all three are below the
$1\times 10^{-6}$ level, this would imply $\hxi{\Bd \to \eta' K^0}\leq
0.09$.

\subsubsection{$\Bu \to \eta K^+$ and $\Bu \to \eta' K^+$.}
\label{sec:bu-to-eta}

The results here update and extend the results
in~\cite{Grossman:2003qp}. As in the previous section, a separate
$\su$ analysis is required for $\Bu \to \eta_1 K^+$ and $\Bu \to
\eta_8 K^+$. The former has a single relation:
\begin{equation}
  \label{eq:43}
  a^q_{\Bu \to \eta_1 K^+} = b^q_{\Bu \to \eta_1 \pi^+}\;.
\end{equation}
For the latter, there are no relations involving just one amplitude. The
relations involving two modes are
\begin{multline}
  \label{eq:44}
  a^q_{\Bu \to \eta_8 K^+} = \frac{1}{\sqrt{3}}\, b^q_{\Bu \to \pi^0
    \pi^+} - \frac{1}{\sqrt{6}}\, b^q_{\Bu \to K^+
    \Kb}\;, \\
  \Longrightarrow\;\;\;\;\;\; \hxi{\Bu \to \eta' K^+} \leq
  0.02\;,\;\;\;\;\; \hxi{\Bu \to \eta K^+} \leq 0.24 \;,
\end{multline}
\begin{multline}
  \label{eq:45}
  a^q_{\Bu \to \eta_8 K^+} = b^q_{\Bu \to \eta_8
    \pi^+} - \sqrt{\frac{2}{3}}\, b^q_{\Bu \to K^+
    \Kb}\;, \\
  \Longrightarrow\;\;\;\;\;\; \hxi{\Bu \to \eta' K^+} \leq
  0.03\;,\;\;\;\;\; \hxi{\Bu \to \eta K^+} \leq 0.50 \;,
\end{multline}
\begin{multline}
  \label{eq:46}
  a^q_{\Bu \to \eta_8 K^+} = \frac{\sqrt{3}}{2}\,b^q_{\Bu \to \pi^0
    \pi^+} - \frac{1}{2}\, b^q_{\Bu \to \eta_8 \pi^+}\;, \\
  \Longrightarrow\;\;\;\;\;\; \hxi{\Bu \to \eta' K^+} \leq
  0.03\;,\;\;\;\;\; \hxi{\Bu \to \eta K^+} \leq 0.43 \;.
\end{multline}
Note that~(\ref{eq:46}) is a linear combination of~(\ref{eq:44})
and~(\ref{eq:45}). 

There are six relations involving three modes but only two of them do
not involve $\Bs$:
\begin{multline}
  \label{eq:47}
  a^q_{\Bu \to \eta_8 K^+} = \frac{1}{\sqrt{6}}\,b^q_{\Bd \to \pi^+
    \pi^-} - \frac{1}{\sqrt{3}}\, b^q_{\Bd \to \pi^0 \pi^0} -
  \frac{1}{\sqrt{6}}\,b^q_{\Bu \to K^+ \Kb}\;, \\
  \Longrightarrow\;\;\;\;\;\; \hxi{\Bu \to \eta' K^+} \leq
  0.02\;,\;\;\;\;\; \hxi{\Bu \to \eta K^+} \leq 0.28 \;,
\end{multline}
\begin{multline}
  \label{eq:48}
  a^q_{\Bu \to \eta_8 K^+} = \frac{1}{2}\sqrt{\frac{3}{2}}\,b^q_{\Bd \to \pi^+
    \pi^-} - \frac{\sqrt{3}}{2}\, b^q_{\Bd \to \pi^0 \pi^0} -
  \frac{1}{2}\,b^q_{\Bu \to \eta_8 \pi^+}\;, \\
  \Longrightarrow\;\;\;\;\;\; \hxi{\Bu \to \eta' K^+} \leq
  0.03\;,\;\;\;\;\; \hxi{\Bu \to \eta K^+} \leq 0.48 \;.
\end{multline}

As was the case in~\cite{Grossman:2003qp}, the strongest constraint on
$\hxi{\Bu \to \eta' K^+}$ (and on $\hxi{\Bu \to \eta K^+}$) comes from
the relation~(\ref{eq:44}) (It is slightly stronger
than~(\ref{eq:47})):
\begin{align}
  \label{eq:49}
  \axi{\Bu \to \eta' K^+} & \leq 0.05\;, \\
  \label{eq:50}
  \axi{\Bu \to \eta K^+} & \leq 0.38\;. 
\end{align}
Note that since $\hxi{\Bu \to \eta' K^+}<\lambda^2$ the upper
bound on $\axi{\Bu \to \eta' K^+}$ in~(\ref{eq:49}), is simply $\lambda^2$.

\subsubsection{$\Bd \to \pi^0 K^0$.}
\label{sec:bd-to-pi0}

The amplitude of the $\Bd \to \pi^0 K^0$ decay is related by $\su$ to
the amplitude of the $b \to s$ transition $\Bd \to \eta_8 K^0$
\begin{equation}
  \label{eq:51}
  a^q_{\Bd \to \pi^0 K^0} = \sqrt{3}\,a^q_{\Bd \to \eta_8 K^0}\;.
\end{equation}
As a consequences, the $\su$ relations can be read off
relations~(\ref{eq:20}),~(\ref{eq:21}),~(\ref{eq:27}) and
(\ref{eq:36})--(\ref{eq:42}). For example, a one amplitude relation is
read off~(\ref{eq:21}):
\begin{equation}
  \label{eq:52}
  a^q_{\Bd \to \pi^0 K^0} = b^q_{\Bs \to \pi^0 \Kb}\;.
\end{equation}
Using the experimental values of $S_{\Bd \to \pi^0 K^0}$, $C_{\Bd \to
  \pi^0 K^0}$ and $\Br{\Bd \to \pi^0 K^0}$ we can solve for the
hadronic parameters and obtain an estimate for the $\Bs \to \pi^0 K_S$
observables:
\begin{align}
  \label{eq:53}
  S_{\Bs \to \pi^0 K_S} & \approx -0.67 \pm 0.27\;, \\
  \label{eq:54}
  C_{\Bs \to \pi^0 K_S} & \approx +0.12 \pm 0.12\;, \\
  \label{eq:55}
  2\,\Br{\Bs \to \pi^0 K_S} & \approx (11 \pm 12)\times 10^{-6}\;.
\end{align}

The single relation involving two amplitudes without $\Bs$ decay is
read off~(\ref{eq:27}):
\begin{equation}
  \label{eq:56}
  a^q_{\Bd \to \pi^0 K^0} = \frac{1}{\sqrt{2}}\, b^q_{\Bd \to K^+ K^-} -
  b^q_{\Bd \to \pi^0 \pi^0}\;.
\end{equation}
This relation gives the strongest constraint
\begin{equation}
  \label{eq:57}
  \hxi{\Bd \to \pi^0 K^0} \leq 0.09\;,
\end{equation}
which results with
\begin{equation}
  \label{eq:58}
  \axi{\Bd \to \pi^0 K^0} \leq 0.15\;.
\end{equation}

The other relations give far weaker constraints.

\subsubsection{$\Bd \to \pi^- K^+$}
\label{sec:bd-to-pi}

A single amplitude relation due to U-spin involves a $\Bs$ mode:
\begin{equation}
  \label{eq:59}
  a^q_{\Bd \to \pi^- K^+}=b^q_{\Bs \to \pi^+ K^-} \;.
\end{equation}

Three amplitude relations involve two modes, but only one does not
involve a $\Bs$ mode:
\begin{multline}
  \label{eq:60}
  a^q_{\Bd \to \pi^- K^+}=b^q_{\Bd \to \pi^+ \pi^-} - b^q_{\Bd \to K^+
    K^-}\;. \;\;\;\;\;\; \Longrightarrow\;\;\;\;\;\; \hxi{\Bd \to \pi^- K^+} \leq 0.12 \;,
\end{multline}
giving
\begin{equation}
  \label{eq:61}
  \axi{\Bd\to \pi^- K^+} \leq 0.19\;.
\end{equation}

There are six relations involving three amplitudes, but only two do
not involve a $\Bs$ mode:
\begin{multline}
  \label{eq:62}
  a^q_{\Bd \to \pi^- K^+} = \sqrt{2}\,b^q_{\Bd \to \pi^0
    \pi^0} -  b^q_{\Bd \to K^- K^+} +
  \sqrt{2}\,b^q_{\Bu \to  \pi^0 \pi^+}\;, \\
  \;\;\;\;\;\; \Longrightarrow\;\;\;\;\;\; \hxi{\Bd \to \pi^- K^+} \leq
  0.27 \;,
\end{multline}
\begin{multline}
  \label{eq:63}
  a^q_{\Bd \to \pi^- K^+} = b^q_{\Bd \to \pi^-
    \pi^+} -  b^q_{\Bd \to K^0 \Kb} -
  \sqrt{3}\,b^q_{\Bu \to  \eta_8 \pi^0}\;, \\
  \;\;\;\;\;\; \Longrightarrow\;\;\;\;\;\; \hxi{\Bd \to \pi^- K^+} \leq
  0.35 \;.
\end{multline}
Clearly (and reasonably), the relation~(\ref{eq:60}) gives the
strongest bound.

\subsubsection{$\Bu \to \pi^+ K^0$}
\label{sec:bu-to-pi+}

A single one amplitude relation due to the U-spin subgroup gives, as
expected, the strongest constraint
\begin{multline}
  \label{eq:64}
  a^q_{\Bu \to \pi^+ K^0} = b^q_{\Bu \to \Kb K^+}\;, 
  \;\;\;\;\;\; \Longrightarrow\;\;\;\;\;\; \hxi{\Bu \to \pi^+ K^0} \leq
  0.05 \;,
\end{multline}
which leads to
\begin{equation}
  \label{eq:65}
  \axi{\Bu \to \pi^+ K^0} \leq 0.05\;.
\end{equation}
In principle, whenever a single amplitude is involved in the
calculation of $\hxi{f}$, a lower bound on $\axi{f}$ can be placed as
well as an upper bound. However, since here $\hxi{\Bu \to \pi^+ K^0}
\sim \lambda^2$ the lower bound is actually zero.

A weaker bound is obtained from the single two amplitudes relation 
\begin{multline}
  \label{eq:66}
  a^q_{\Bu \to \pi^+ K^0} = \sqrt{\frac{3}{2}}\,b^q_{\Bu \to \eta_8
    \pi^+} -  \sqrt{\frac{1}{2}}\,b^q_{\Bu \to \pi^0 \pi^+}\;,
  \;\;\;\;\;\; \Longrightarrow\;\;\;\;\;\; \hxi{\Bu \to \pi^+ K^0}
  \leq 0.21 \;.
\end{multline}

Three relations involve three amplitudes but only one does not involve
a $\Bs$ mode
\begin{multline}
  \label{eq:67}
  a^q_{\Bu \to \pi^+ K^0} = \sqrt{\frac{3}{2}}\,b^q_{\Bu \to \eta_8
    \pi^+} +  \sqrt{\frac{1}{2}}\,b^q_{\Bd \to \pi^0 \pi^0} -
  \frac{1}{2}\,b^q_{\Bd \to \pi^- \pi^+} \;, \\
  \;\;\;\;\;\; \Longrightarrow\;\;\;\;\;\; \hxi{\Bu \to \pi^+ K^0}
  \leq 0.23 \;.
\end{multline}
This relation gives a still weaker bound.

\subsubsection{$\Bu \to \pi^0 K^+$}
\label{sec:bd-to-pi0-1}

Three amplitude relations involve two modes
\begin{multline}
  \label{eq:68}
  a^q_{\Bu \to \pi^0 K^+} = b^q_{\Bu \to \pi^0
    \pi^+} +  \sqrt{\frac{1}{2}}\,b^q_{\Bu \to \Kb K^+}\;,
  \;\;\;\;\;\; \Longrightarrow\;\;\;\;\;\; \hxi{\Bu \to \pi^0 K^+}
  \leq 0.20 \;,
\end{multline}
\begin{multline}
  \label{eq:69}
  a^q_{\Bu \to \pi^0 K^+} = \sqrt{3}\,b^q_{\Bu \to \eta_8
    \pi^+} -  \sqrt{\frac{1}{2}}\,b^q_{\Bu \to \Kb K^+}\;,
  \;\;\;\;\;\; \Longrightarrow\;\;\;\;\;\; \hxi{\Bu \to \pi^0 K^+}
  \leq 0.32 \;,
\end{multline}
\begin{multline}
  \label{eq:70}
  a^q_{\Bu \to \pi^0 K^+} = \frac{1}{2}\,b^q_{\Bu \to \pi^0
    \pi^+} +  \sqrt{\frac{3}{4}}\,b^q_{\Bu \to \eta_8 \pi^+}\;,
  \;\;\;\;\;\; \Longrightarrow\;\;\;\;\;\; \hxi{\Bu \to \pi^0 K^+}
  \leq 0.21 \;.
\end{multline}
The relation~(\ref{eq:70}) is a linear combination of~(\ref{eq:68})
and~(\ref{eq:69}). Two out of six relations involving three amplitudes
do not involve $\Bs$ modes
\begin{multline}
  \label{eq:71}
  a^q_{\Bu \to \pi^0 K^+} = \sqrt{\frac{1}{2}}\,b^q_{\Bd \to \pi^-
    \pi^+} - b^q_{\Bd \to \pi^0\pi^0} +  \sqrt{\frac{1}{2}}\,b^q_{\Bu
    \to \Kb K^+}\;, \\
  \;\;\;\;\;\; \Longrightarrow\;\;\;\;\;\; \hxi{\Bu \to \pi^0 K^+}
  \leq 0.23 \;,
\end{multline}
\begin{multline}
  \label{eq:72}
  a^q_{\Bu \to \pi^0 K^+} = \sqrt{\frac{1}{8}}\,b^q_{\Bd \to \pi^-
    \pi^+} -  \frac{1}{2} \,b^q_{\Bd  \to \pi^0\pi^0} +
  \sqrt{\frac{3}{4}}\,b^q_{\Bu \to \eta_8 \pi^+}\;, \\ 
  \;\;\;\;\;\; \Longrightarrow\;\;\;\;\;\; \hxi{\Bu \to \pi^0 K^+}
  \leq 0.23 \;.
\end{multline}
The strongest bound~(\ref{eq:68}) implies
\begin{equation}
  \label{eq:73}
  \axi{\Bu \to \pi^0 K^+} \leq 0.31\;.
\end{equation}

\subsection{$B \to VP$ modes.}
\label{sec:b-to-vp}

The are sixteen $b \to s$ decay processes in which one final meson
belongs to the vector nonet and the other belongs to the pseudo-scalar
nonet. The method of calculation is generally similar to the $B \to
PP$ case but using the appropriate tables of reduced matrix elements
(with table~\ref{tab:B-to-P_8V_8} taking the role of
table~\ref{tab:B-to-M_8N_8}). We consider the relevant amplitude
relations and the resulting constraints below

\subsubsection{$B \to \phi K^0$ and $B \to \omega K^0$}
\label{sec:b-to-omega}

The $\omega - \phi$ mixing is given by~\cite{Eidelman:2004wy}
\begin{equation}
  \label{eq:74}
    \begin{aligned}
    \omega &= \sqrt{\frac{2}{3}}\, \phi_1 +
    \sqrt{\frac{1}{3}}\,\phi_8\;, \\ 
    \phi &= \sqrt{\frac{1}{3}}\, \phi_1 -
    \sqrt{\frac{2}{3}}\, \phi_8 \;. 
  \end{aligned}
\end{equation}

We need to discuss both $B \to \phi_1 K^0$ and $B \to \phi_8
K^0$. Since, similarly to $\eta_1$, $\phi_1$ is a singlet of $\su$,
the same $\su$ relations obtained from table~\ref{tab:B-to-S_1M_8}
hold. We can therefore simply adapt the single relation~(\ref{eq:19}):
\begin{equation}
  \label{eq:75}
    a^q_{\Bd \to \phi_1 K^0} = \sqrt{\frac{3}{2}}\,b^q_{\Bd \to \phi_1
    \eta_8}- \sqrt{\frac{1}{2}}\,b^q_{\Bd \to \phi_1 \pi^0}\;.
\end{equation}

Turning now to the $\Bd \to \phi_8 K^0$ mode, the single relation
involving two amplitudes and five out of the six relations involving
three amplitudes, involve $\Bs$ modes. The remaining single amplitude
with three amplitudes which does not involve $\Bs$ modes is
\begin{equation}
  \label{eq:76}
  a^q_{\Bd \to \phi_8 K^0} = \sqrt{\frac{3}{2}}\, b^q_{\Bd \to 
    \phi_8 \eta_8} -\sqrt{\frac{3}{2}}\, b^q_{\Bd \to \Ksb K^0}
  - \sqrt{\frac{1}{2}}\,b^q_{\Bd \to \phi_8 \pi^0 }\;.
\end{equation}

Combining~(\ref{eq:75}) and~(\ref{eq:76}) we obtain
\begin{align}
  \begin{split}
    a^q_{\Bd \to \phi K^0} &= b^q_{\Bd \to  \Ksb K^0} -
    \sqrt{\frac{1}{2}}\,b^q_{\Bd \to \phi \pi^0} +
    \sqrt{\frac{3}{2}}\,\cos \theta_{\eta\eta'}\,b^q_{\phi \eta} -
    \sqrt{\frac{3}{2}}\,\sin \theta_{\eta\eta'}\,b^q_{\phi \eta'} \;,
  \end{split} \label{eq:77} \\
  \begin{split}
     a^q_{\Bd \to \omega K^0} &= -\sqrt{\frac{1}{2}}\,b^q_{\Bd \to  \Ksb K^0} -
    \sqrt{\frac{1}{2}}\,b^q_{\Bd \to \omega \pi^0} +
    \sqrt{\frac{3}{2}}\,\cos \theta_{\eta\eta'}\,b^q_{\omega \eta} -
    \sqrt{\frac{3}{2}}\,\sin \theta_{\eta\eta'}\,b^q_{\omega \eta'} \;.
  \end{split} \label{eq:78}
\end{align}
Unfortunately, a constraint on $\Br{\Bd \to \Ksb K^0}$ is currently
not available. As a consequence, a constraint on $\axi{\Bd \to \phi
  K^0}$ and $\axi{\Bd \to \phi K^0}$ cannot be obtained. In fact, it
can be shown that any amplitude relation for $\Bd \to \phi_8 K^0$ must
involve at least one of the modes $\Bd \to K^{*+}K^-$, $\Bd \to
K^{*-}K^+$, $\Bd \to \Ksb K^0$ or $\Bd \to K^{*0} \Kb$. As long as the
branching ratios for these modes remain unknown, no bound can be
obtained.

We can, however, write $\hxi{\Bd \to \phi K^0}$ and $\hxi{\Bd \to
  \omega K^0}$ as a function of the missing branching ratio. We get
\begin{align}
  \label{eq:79}
  \hxi{\Bd \to \phi K^0} & \leq 0.21 + 0.08\,\sqrt{10^6 \times \Br{\Bd
      \to \Ksb K^0}} \;, \\
  \label{eq:80}
  \hxi{\Bd \to \omega K^0} & \leq 0.31 + 0.07\,\sqrt{10^6 \times \Br{\Bd
      \to \Ksb K^0}} \;.
\end{align}

\subsubsection{$\Bu \to \phi K^+$ and $\Bu \to \omega K^+$}
\label{sec:bu-to-phi}

The relation for $\Bu \to \phi_1 K^+$ can be read
off~(\ref{eq:43}). There is a single amplitude relation
\begin{equation}
  \label{eq:81}
  a^q_{\Bu \to \phi_1 K^+} = b^q_{\Bu \to \phi_1 \pi^+}\;.
\end{equation}

For $\Bu \to \phi_8 K^+$ we find three relations involving two
amplitudes:
\begin{multline}
  \label{eq:82}
  a^q_{\Bu \to \phi_8 K^+} = \sqrt{\frac{1}{3}}\,b^q_{\Bu \to \rho^0
    \pi^+} - \sqrt{\frac{1}{6}}\,b^q_{\Bu  \to \Ksb K^+}\;, \\
  \;\;\;\;\;\; \Longrightarrow\;\;\;\;\;\; \hxi{\Bu \to \phi K^+}
  \leq 0.26 \;,\;\;\;\;\; \hxi{\Bu \to \omega K^+} \leq 0.31\;,
\end{multline}
\begin{multline}
  \label{eq:83}
  a^q_{\Bu \to \phi_8 K^+} = b^q_{\Bu \to \phi_8
    \pi^+} - \sqrt{\frac{3}{2}}\,b^q_{\Bu  \to \Ksb K^+}\;, \\
  \;\;\;\;\;\; \Longrightarrow\;\;\;\;\;\; \hxi{\Bu \to \phi K^+}
  \leq 0.22 \;,\;\;\;\;\; \hxi{\Bu \to \omega K^+} \leq 0.36\;,
\end{multline}
\begin{multline}
  \label{eq:84}
  a^q_{\Bu \to \phi_8 K^+} = \sqrt{\frac{3}{4}}\,b^q_{\Bu \to \rho^0
    \pi^+} - \frac{1}{2}\,b^q_{\Bu  \to \phi_8 \pi^+}\;, \\
  \;\;\;\;\;\; \Longrightarrow\;\;\;\;\;\; \hxi{\Bu \to \phi K^+}
  \leq 0.29 \;,\;\;\;\;\; \hxi{\Bu \to \omega K^+} \leq 0.28\;,
\end{multline}
The relation~(\ref{eq:84}) is a linear combination
of~(\ref{eq:82}) and~(\ref{eq:83}). Those relations are similar to
relations~(\ref{eq:44})--(\ref{eq:46}).

As was the case in~\cite{Grossman:2003qp}, relation~(\ref{eq:83})
gives the strongest bound on $\axi{\Bu \to \phi K^+}$ which currently
implies
\begin{equation}
  \label{eq:85}
  \axi{\Bu \to \phi K^+} \leq 0.34\;. 
\end{equation}
Relation~(\ref{eq:84}) is the one that gives the strongest bound on
$\axi{\Bu \to \omega K^+}$ which implies
\begin{equation}
  \label{eq:86}
  \axi{\Bu \to \omega K^+} \leq 0.46 \;.
\end{equation}
There are no amplitude relations involving three modes.

\subsubsection{$\Bd \to \rho^0 K^0$}
\label{sec:bd-to-rho0}

The single relation that involves two amplitudes, involves $\Bs$
modes. Out of the six amplitude relation that involves three
amplitude, only one does not involve $\Bs$ modes:
\begin{equation}
  \label{eq:87}
  a^q_{\Bd \to \rho^0 K^0} = \sqrt{\frac{1}{2}}\,b^q_{\Bd \to \Ksb
    K^0} + \sqrt{\frac{3}{2}}\,b^q_{\Bd \to \rho^0 \eta_8}  
  -\sqrt{\frac{1}{2}}\,b^q_{\Bd \to \rho^0 \pi^0}\;.
\end{equation}
Similarly to $\Bd \to \phi K^0$ and $\Bd \to \omega K^0$, the fact
that the branching ratio $\Br{\Bd \to \Ksb K^0}$ has not been measured
yet, prevents a bound from being obtained. At least one of the four
modes $\Bd \to K^{*+}K^-$, $\Bd \to K^{*-}K^+$, $\Bd \to \Ksb K^0$ or
$\Bd \to K^{*0} \Kb$, must be measured in order for a bound to be
placed.

We can write the resulting constraint as a function of the
missing branching ratio:
\begin{equation}
  \label{eq:88}
  \hxi{\Bd \to \rho^0 K^0} \leq 0.32 + 0.07\,\sqrt{10^6 \times \Br{\Bd
      \to \Ksb K^0}}\;.
\end{equation}

\subsubsection{$\Bd \to \rho^- K^+$}
\label{sec:bd-to-rho}

We note first a single amplitude U-spin relation
\begin{equation}
  \label{eq:89}
  a^q_{\Bd \to \rho^- K^+} = b^q_{\Bs \to K^{*-} \pi^+}\;.
\end{equation}

On a more practical ground, we find a single amplitude relation
involving two modes
\begin{equation}
  \label{eq:90}
  a^q_{\Bd \to \rho^- K^+} = b^q_{\Bd \to \rho^- \pi^+} -b^q_{\Bd \to
    K^{*-} K^+}\;.
\end{equation}
$\Br{\Bd \to K^{*-} K^+}$ is needed in order to place a bound from
eq.~\eqref{eq:90}, or, more generally, at least one of the four modes $\Bd \to
K^{*+}K^-$, $\Bd \to K^{*-}K^+$, $\Bd \to \Ksb K^0$ or $\Bd \to K^{*0}
\Kb$ is needed.

As a function of the missing branching ratio we get
\begin{equation}
  \label{eq:91}
  \hxi{\Bd \to \rho^- K^+} \leq 0.34 + 0.07\,\sqrt{10^6 \times \Br{\Bd
      \to K^{*-} K^+}}\;.
\end{equation}

\subsubsection{$\Bu \to \rho^0 K^+$}
\label{sec:bu-to-rho0}

Three amplitude relations involve two modes
\begin{multline}
  \label{eq:92}
  a^q_{\Bu \to \rho^0 K^+} = \sqrt{\frac{1}{2}}\,b^q_{\Bu \to \Ksb
    K^+} + b^q_{\Bu  \to \rho^0 \pi^+}\;,
  \;\;\;\;\;\; \Longrightarrow\;\;\;\;\;\; \hxi{\Bu \to \rho^0 K^+}
  \leq 0.49 \;,
\end{multline}
\begin{multline}
  \label{eq:93}
  a^q_{\Bu \to \rho^0 K^+} = - \sqrt{\frac{1}{2}}\,b^q_{\Bu \to \Ksb
    K^+} + \sqrt{3}\,b^q_{\Bu  \to \phi_8 \pi^+}\;,
  \;\;\;\;\;\; \Longrightarrow\;\;\;\;\;\; \hxi{\Bu \to \rho^0 K^+}
  \leq 0.55 \;,
\end{multline}
\begin{multline}
  \label{eq:94}
  a^q_{\Bu \to \rho^0 K^+} = \frac{1}{2}\,b^q_{\Bu \to \rho^0
    \pi^+} + \sqrt{\frac{3}{4}}\,b^q_{\Bu  \to \phi_8 \pi^+}\;,
  \;\;\;\;\;\; \Longrightarrow\;\;\;\;\;\; \hxi{\Bu \to \rho^0 K^+}
  \leq 0.34 \;.
\end{multline}
Relation~(\ref{eq:94}) is a linear combination of~(\ref{eq:92})
and~(\ref{eq:93}). The strongest bound~(\ref{eq:94}) implies
\begin{equation}
  \label{eq:95}
  \axi{\Bu \to \rho^0 K^+} \leq 0.61\;.
\end{equation}

\subsubsection{$\Bu \to \rho^+ K^0$}
\label{sec:bu-to-rho+}

The branching ratio $\Br{\Bu \to \rho^+ K^0}$ has currently only an
upper bound. As a consequence, no bound on $\axi{\Bu \to \rho^+ K^0}$
can be placed. We list here the amplitude relations that will become
useful once this branching ratio is measured.

A single one amplitude relation is
\begin{equation}
  \label{eq:96}
  a^q_{\Bu \to \rho^+ K^0} = b^q_{\Bu \to K^{*+} \Kb}\;.
\end{equation}
The branching ratio $\Br{\Bu \to K^{*+} \Kb}$ has not been measured
yet.

There is a single relation involving two modes:
\begin{multline}
  \label{eq:97}
  a^q_{\Bu \to \rho^+ K^0} = \sqrt{\frac{3}{2}}\,b^q_{\Bu \to \rho^+
    \eta_8} - \sqrt{\frac{1}{2}}\,b^q_{\Bu \to \rho^+ \pi^0}\;,\\
  \;\;\;\;\;\; \Longrightarrow\;\;\;\;\;\; \hxi{\Bu \to \rho^+ K^0}
  \leq \frac{1.66}{\sqrt{10^6 \times \Br{\Bu \to \rho^+ K^0}}} \;.
\end{multline}

There are no amplitude relations involving three modes.

Considering the upper bound on $\Br{\Bd \to \rho^+ K^0}$, a strong
bound can still result.

\subsubsection{$\Bd \to K^{*0} \eta$ and $\Bd \to K^{*0} \eta'$}
\label{sec:bd-to-k0}

The branching ratio $\Br{\Bd \to K^{*0} \eta'}$ has currently only
been bounded. As a consequence, a bound on $\axi{\Bd \to K^{*0}
  \eta'}$ cannot be placed.

The relation for $\Bd \to K^{*0} \eta_1$ can be read
off~(\ref{eq:19}):
\begin{equation}
  \label{eq:98}
  a^q_{\Bd \to K^{*0} \eta_1} = \sqrt{\frac{3}{2}}\,b^q_{\Bd \to
    \phi_8 \eta_1}- \sqrt{\frac{1}{2}}\,b^q_{\Bd \to \rho^0 \eta_1}\;.
\end{equation}
A single relation for $\Bd \to K^{*0} \eta_8$ involving two
amplitudes, and five out of six relation involving three amplitudes,
involve $\Bs$ modes. The one additional relation involving three
amplitude is
\begin{equation}
  \label{eq:99}
  a^q_{\Bd \to K^{*0} \eta_8} = \sqrt{\frac{3}{2}}\, b^q_{\Bd \to 
    \phi_8 \eta_8} -\sqrt{\frac{3}{2}}\, b^q_{\Bd \to K^{*0} \Kb}
  - \sqrt{\frac{1}{2}}\,b^q_{\Bd \to \rho^0 \eta_8}\;.
\end{equation}
Since $\Br{\Bd \to K^{*0} \Kb}$ has not been constrained yet,
no bound on $\axi{\Bd \to K^{*0} \eta}$ and  $\axi{\Bd \to K^{*0}
  \eta'}$ can be currently obtained. At least one of the four $\Bd \to
K^{*}K$ modes is required for a bound.

We can write the resulting constraint as a function of the missing
branching ratio. Since $\Br{\Bd \to K^{*0} \eta'}$ has not been
measured we only have
\begin{equation}
  \label{eq:100}
    \hxi{\Bd \to K^{*0} \eta} \leq 0.14 + 0.06\,\sqrt{10^6 \times \Br{\Bd
      \to K^{*0} \Kb}}\;.
\end{equation}

\subsubsection{$\Bu \to K^{*+} \eta$ and $\Bu \to K^{*+} \eta'$}
\label{sec:bu-to-k+}

$\Br{\Bu \to K^{*+} \eta'}$ has not been measured yet and therefore
$\axi{\Bu \to K^{*+} \eta'}$ cannot be bounded. The single relation
for $\Bu \to K^{*+} \eta_1$ can be read off~(\ref{eq:43}):
\begin{equation}
  \label{eq:101}
  a^q_{\Bu \to K^{*+} \eta_1} = b^q_{\Bu \to \rho^+ \eta_1}\;.
\end{equation}
For $\Bu \to K^{*+} \eta_8$ the relations are equivalent
to~(\ref{eq:44})--(\ref{eq:46}):
\begin{align}
  \label{eq:102}
  a^q_{\Bu \to  K^{*+} \eta_8} & = \frac{1}{\sqrt{3}}\, b^q_{\Bu \to
    \rho^+ \pi^0} - \frac{1}{\sqrt{6}}\, b^q_{\Bd \to K^{*+} \Kb}\;,
  \\ 
  \label{eq:103}
  a^q_{\Bu \to K^{*+} \eta_8} & = b^q_{\Bu \to \rho^+ \eta_8} -
  \sqrt{\frac{2}{3}}\, b^q_{\Bd \to K^{*+} \Kb}\;, \\
  \label{eq:104}
  a^q_{\Bu \to K^{*+} \eta_8} & = \sqrt{\frac{3}{4}}\,b^q_{\Bu \to
    \rho^+ \pi^0} - \frac{1}{2}\, b^q_{\Bu \to \rho^+ \eta_8}\;.
\end{align}
Relation~(\ref{eq:104}) is a linear combination of~(\ref{eq:102})
and~(\ref{eq:103}).

Since $\Br{\Bu \to K^{*+} \Kb}$ has not been bounded yet, the only
useful relation currently is~(\ref{eq:104}). We get
\begin{equation}
  \label{eq:105}
  \hxi{\Bu \to K^{*+} \eta} \leq 0.26\;,\;\;\;\;\;\;\;\;\;\;\;\;\; \hxi{\Bu \to
    K^{*+} \eta'} \leq \frac{1.37}{\sqrt{10^6 \times \Br{\Bu \to K^{*+} \eta'}}} \;,
\end{equation}
which gives
\begin{equation}
  \label{eq:106}
  \axi{\Bu \to K^{*+} \eta} \leq 0.42\;.
\end{equation}
At the current upper bound of $\Br{\Bu \to K^{*+} \eta'}$ the
constraint on $\hxi{\Bu \to K^{*+} \eta'}$ is already not very strong.

\subsubsection{$\Bd \to K^{*0} \pi^0$}
\label{sec:bd-to-k0-1}

There is one amplitude relation involving two modes and six amplitude
relations involving three modes. The only relation, however, which
does not involve $\Bs$ modes is
\begin{equation}
  \label{eq:107}
  a^q_{\Bd \to K^{*0} \pi^0} = \sqrt{\frac{1}{2}}\,b^q_{\Bd \to K^{*0}
    \Kb} -\sqrt{\frac{1}{2}}\,b^q_{\rho^0 \pi^0} +
  \sqrt{\frac{3}{2}}\,b^q_{\Bd \to \phi_8 \pi^0}\;.
\end{equation}
Once more, since $\Br{\Bd \to K^{*0} \Kb}$ is not bounded, no bound on
$\axi{\Bd \to K^{*0} \Kb}$ is attained. Again, at least one of the
four $\Bd \to K^{*}K$ already mentioned must be bounded in order for
$\axi{\Bd \to K^{*0} \Kb}$ to be bounded.

As a function of the missing branching ratio we can write
\begin{equation}
  \label{eq:108}
  \hxi{\Bd \to K^{*0} \pi^0} \leq 0.46 + 0.12\,\sqrt{10^6 \times \Br{\Bd
      \to K^{*0} \Kb}}\;,
\end{equation}
which seems to give a rather weak bound in any case.

\subsubsection{$\Bd \to K^{*+} \pi^-$}
\label{sec:bd-to-k+}

We first note a U-spin relation involving one amplitude (with a $\Bs$ mode):
\begin{equation}
  \label{eq:109}
  a^q_{\Bd \to K^{*+} \pi^-} = b^q_{\Bs \to \rho^+ K^-}\;.
\end{equation}

There is a single relation involving two amplitudes
\begin{equation}
  \label{eq:110}
  a^q_{\Bd \to K^{*+} \pi^-} = b^q_{\Bd \to \rho^+ \pi^-} - b^q_{\Bd
    \to K^{*+} K^-}\;.
\end{equation}
There is no relation involving three modes (or four). At least
one of the four $\Bd \to K^{*}K$ is needed. Since $\Br{\Bd \to K^{*+}
  K^-}$ is not bounded, $\axi{\Bd \to K^{*+} \pi^-}$ is not bounded
currently.

As a function of the missing branching ratio we can write
\begin{equation}
  \label{eq:111}
  \hxi{\Bd \to K^{*+} \pi^-} \leq 0.32 + 0.06\,\sqrt{10^6 \times \Br{\Bd
      \to K^{*+} K^-}}\;.
\end{equation}

\subsubsection{$\Bu \to K^{*+} \pi^0$}
\label{sec:bu-to-k+-1}

Three amplitude relations involve two modes (similar
to~(\ref{eq:92})--(\ref{eq:94})):
\begin{align}
  \label{eq:112}
  a^q_{\Bu \to K^{*+} \pi^0} & = \sqrt{\frac{1}{2}}\,b^q_{\Bu \to K^{*+}
    \Kb} + b^q_{\Bu  \to \rho^+ \pi^0}\;,\\
  \label{eq:113}
  a^q_{\Bu \to K^{*+} \pi^0} & = - \sqrt{\frac{1}{2}}\,b^q_{\Bu \to K^{*+}
    \Kb} + \sqrt{3}\,b^q_{\Bu  \to \rho^+ \eta_8}\;,\\
  \label{eq:114}
  a^q_{\Bu \to K^{*+} \pi^0} & = \frac{1}{2}\,b^q_{\Bu \to \rho^+
    \pi^0} + \sqrt{\frac{3}{4}}\,b^q_{\Bu  \to \rho^+ \eta_8}\;.
\end{align}
Relation~(\ref{eq:114}) is a linear combination of~(\ref{eq:112})
and~(\ref{eq:113}). 

Since $\Br{\Bu \to K^{*+} \Kb}$ has not been bounded yet a 
bound comes only from~(\ref{eq:114}):
\begin{equation}
  \label{eq:115}
  \hxi{\Bu \to K^{*+} \pi^0} \leq 0.45\;,
\end{equation}
which implies
\begin{equation}
  \label{eq:116}
  \axi{\Bu \to K^{*+} \pi^0} \leq 0.91\;,
\end{equation}

\subsubsection{$\Bu \to K^{*0} \pi^+$}
\label{sec:bu-to-k0}

A single U-spin relation involving one amplitude gives
\begin{multline}
  \label{eq:117}
  a^q_{\Bu \to K^{*0} \pi^+} =  b^q_{\Bu \to \Ksb K^+}\;, 
  \;\;\;\;\;\; \Longrightarrow\;\;\;\;\;\; \hxi{\Bu \to K^{*0} \pi^+} \leq
  0.16 \;.
\end{multline}

A single relation involving two amplitudes is
\begin{multline}
  \label{eq:118}
  a^q_{\Bu \to K^{*0} \pi^+} = \sqrt{\frac{3}{2}}\,b^q_{\Bu \to \phi_8
    \pi^+} -  \sqrt{\frac{1}{2}}\,b^q_{\Bu \to \rho^0 \pi^+}\;,
  \;\;\;\;\;\; \Longrightarrow\;\;\;\;\;\; \hxi{\Bu \to K^{*0} \pi^+}
  \leq 0.30 \;.
\end{multline}
There are no other relations.

As expected, the strongest bound is obtained from~(\ref{eq:117})
\begin{equation}
  \label{eq:119}
  \axi{\Bu \to K^{*0} \pi^+}  \leq 0.25 \;.
\end{equation}

\subsection{$B \to VV$ modes}
\label{sec:b-to-vv-2}

The relations for $b \to s$ transitions in $B \to VV$ decays can be
read off the $B \to PP$ relations by a direct substitution. The only
additional difference is that the $\eta$--$\eta'$ mixing should be
replaced by the $\omega$--$\phi$ mixing with a different mixing angle.

Currently, only four $b \to s$ $B \to VV$ are measured (the
other four are only bounded). We present below the details.

\subsubsection{$\Bd \to \phi K^{*0}$ and $\Bd \to \omega K^{*0}$}
\label{sec:bd-to-phi}

The relevant relations are those that correspond to the relations in
section~\ref{sec:bd-to-eta-1}. However, the relation equivalent
to~(\ref{eq:19}) involves the mode $\Bd \to \phi_1 \phi_8$ which, by
mixing, involves $\Br{\Bd \to \omega \omega}$, $\Br{\Bd \to \phi
  \omega}$ and $\Br{\Bd \to \phi \phi}$. Only the latter have been
currently bounded and therefore no bounds on $\axi{\Bd \to \phi
  K^{*0}}$ or $\axi{\Bd \to \omega K^{*0}}$ result.

We can express the constraint as a function of the missing branching
ratios. Since only $\Br{\Bd \to \phi K^{*0}}$ is currently measured,
we only consider this mode. The most useful relation is the one
corresponding to relations~\eqref{eq:42} which gives:
\begin{equation}
  \label{eq:120}
  \hxi{\Bd \to \phi K^{*0}} \leq 0.15 + 0.03\,\sqrt{10^6 \times
    \Br{\Bd \to \omega \omega}} \;, \\
\end{equation}

\subsubsection{$\Bu \to \phi K^{*+}$ and $\Bu \to \omega K^{*0}$}
\label{sec:bu-to-phi-1}

The relevant relations are those corresponding to the ones in
section~\ref{sec:bu-to-eta}. Due to the different mixing, the
strongest bounds on $\axi{\Bu \to \phi K^{*+}}$ and $\axi{\Bu \to
  \omega K^{*+}}$ both come from the relation which corresponds
to~(\ref{eq:48}). Since $\Br{\Bu \to \omega K^{*+}}$ has only been
bounded, we get
\begin{equation}
  \label{eq:121}
  \hxi{\Bu \to \omega K^{*+}} \leq \frac{1.53}{\sqrt{10^6 \times
      \Br{\Bu \to \omega K^{*+}}}} \;. 
\end{equation}
Already at the current upper bound of $\Br{\Bu \to \omega K^{*+}}$ a
useful bound on $\axi{\Bu \to \omega K^{*+}}$ does not result.

For $\Bu \to \phi K^{*+}$ we get
\begin{equation}
  \label{eq:122}
    \hxi{\Bu \to \phi K^{*+}} \leq 0.41\;,
\end{equation}
which leads to
\begin{equation}
  \label{eq:123}
    \axi{\Bu \to \phi K^{*+}} \leq 0.78\;.
\end{equation}

\subsubsection{$\Bd \to \rho^0 K^{*0}$}
\label{sec:bd-to-rho0-1}

The relevant relation correspond to the ones in
section~\ref{sec:bd-to-pi0}:
\begin{equation}
  \label{eq:124}
  a^q_{\Bd \to \rho^0 K^{*0}} = \sqrt{3}\,b^q_{\Bd \to \phi_8 K^{*0}}
\end{equation}
the actual relation can be therefore read off the ones in
section~\ref{sec:bd-to-eta-1}.

Since $\Br{\Bd \to \rho^0 K^{*0}}$ is not measured yet,
we cannot give an explicit bound here. Had the branching ratio been
measured, the strongest bound would come from the relation
corresponding to~(\ref{eq:38}):
\begin{equation}
  \label{eq:125}
  \hxi{\Bd \to \rho^0 K^{*0}} \leq \frac{2.04}{\sqrt{10^6 \times
      \Br{\Bd \to \rho^0 K^{*0}}}}\;. 
\end{equation}
Already at the current upper bound of $\Br{\Bd \to \rho^0 K^{*0}}$ a
useful bound does not result.

\subsubsection{$\Bd \to \rho^- K^{*+}$}
\label{sec:bd-to-rho-1}

The relevant relations correspond to those in
section~\ref{sec:bd-to-pi}. Again, since $\Br{\Bd \to \rho^- K^{*+}}$
is not measured yet, no bound on $\axi{\Bd \to \rho^-
  K^{*+}}$ can be attained. Had it been measured, the strongest bound
would have come from a relation corresponding to~(\ref{eq:63})
\begin{equation}
  \label{eq:126}
  \hxi{\Bd \to \rho^- K^{*+}} \leq \frac{3.68}{\sqrt{10^6 \times
      \Br{\Bd \to \rho^- K^{*+}}}}\;. 
\end{equation}

\subsubsection{$\Bu \to \rho^+ K^{*0}$}
\label{sec:bu-to-rho+-1}

The relevant relations are those corresponding to the ones in
section~\ref{sec:bu-to-pi+}. The strongest bound comes from a relation
corresponding to~(\ref{eq:64}):
\begin{equation}
  \label{eq:127}
  \axi{\Bu \to \rho^+ K^{*0}} \leq 1.44\;.
\end{equation}

\subsubsection{$\Bu \to \rho^0 K^{*+}$}
\label{sec:bu-to-rho0-1}

The relevant relations corresponds to those in
section~\ref{sec:bd-to-pi0-1}. The strongest bound comes from the
relation corresponding to~(\ref{eq:70}) which gives
\begin{equation}
  \label{eq:128}
  \axi{\Bu \to \rho^0 K^{*+}} \leq 1.02\;.
\end{equation}

\section{Summary}
\label{sec:summary}

We summarize the constrains on the various $\axi{B \to f}$'s obtained
in section~\ref{sec:results}. For $B \to PP$ modes, the only mode
which is not measured yet is $\Bd \to \eta K^0$. For the other modes
we get the following bounds
\begin{align}
  \axi{\Bd \to \eta' K^0} & \leq 0.26\;, \tag{\ref{eq:31}} \\
  \axi{\Bd \to \pi^0 K^0} & \leq 0.15\;, \tag{\ref{eq:58}} \\
  \axi{\Bd \to \pi^- K^+} & \leq 0.19 \;, \tag{\ref{eq:61}} 
\end{align} 
\begin{align}
  \axi{\Bu \to \eta' K^+} & \leq 0.05 \;, \tag{\ref{eq:49}} \\
  \axi{\Bu \to \eta K^+} & \leq 0.38 \;, \tag{\ref{eq:50}} \\
  \axi{\Bu \to \pi^+ K^0} & \leq 0.05 \;, \tag{\ref{eq:65}} \\
  \axi{\Bu \to \pi^0 K^+} & \leq 0.31 \;. \tag{\ref{eq:73}}
\end{align}

For $B \to VP$ modes we have three modes which have not been measured
yet and have only an upper bound: $\Bd \to K^{*0} \eta'$, $\Bu \to
\rho^+ K^0$ and $\Bu \to K^{*+} \eta'$. The $\Bd \to VP$ modes are
unbounded due to the lack of bounds on various $\Bd \to K^*K$
modes. We therefore write $\hxi{}$ for these modes as a function of
the missing branching ratios. We get the following results:
\begin{align}
    \hxi{\Bd \to \phi K^0} & \leq 0.21 + 0.08\,\sqrt{10^6 \times \Br{\Bd
      \to \Ksb K^0}} \;, \tag{\ref{eq:79}} \\
 \hxi{\Bd \to \omega K^0} & \leq 0.31 + 0.07\,\sqrt{10^6 \times \Br{\Bd
      \to \Ksb K^0}} \;, \tag{\ref{eq:80}} \\
  \hxi{\Bd \to \rho^0 K^0} & \leq 0.32 + 0.07\,\sqrt{10^6 \times \Br{\Bd
      \to \Ksb K^0}}\;,  \tag{\ref{eq:88}} \\
  \hxi{\Bd \to \rho^- K^+} & \leq 0.34 + 0.07\,\sqrt{10^6 \times \Br{\Bd
      \to K^{*-} K^+}}\;,  \tag{\ref{eq:91}} \\
  \hxi{\Bd \to K^{*0} \eta} & \leq 0.14 + 0.06\,\sqrt{10^6 \times \Br{\Bd
      \to K^{*0} \Kb}}\;,  \tag{\ref{eq:100}} \\
  \hxi{\Bd \to K^{*0} \pi^0} & \leq 0.46 + 0.12\,\sqrt{10^6 \times \Br{\Bd
      \to K^{*0} \Kb}}\;,  \tag{\ref{eq:108}} \\
  \hxi{\Bd \to K^{*+} \pi^-} & \leq 0.32 + 0.06\,\sqrt{10^6 \times \Br{\Bd
      \to K^{*+} K^-}}\;,  \tag{\ref{eq:111}} 
\end{align}
\begin{align}
  \axi{\Bu \to \phi K^+} & \leq 0.34 \;, 
  \tag{\ref{eq:85}} \\
  \axi{\Bu \to \omega K^+} & \leq 0.46 \;, 
  \tag{\ref{eq:86}} \\
  \axi{\Bu \to \rho^0 K^+} & \leq 0.61 \;, 
  \tag{\ref{eq:95}} \\
  \axi{\Bu \to K^{*+} \eta} & \leq 0.42 \;,
  \tag{\ref{eq:106}} \\
  \axi{\Bu \to K^{*+} \pi^0} & \leq 0.91 \;,
  \tag{\ref{eq:116}} \\
  \axi{\Bu \to K^{*0} \pi^+} & \leq 0.25 \;.
  \tag{\ref{eq:119}}
\end{align}

For $B \to VV$ modes there are four modes which are still to be
measured. Those are $\Bd \to \omega K^{*0}$, $\Bd
\to \rho^0 K^{*0}$, $\Bd \to \rho^- K^{*+}$ and $\Bu \to \omega
K^{*+}$. For the other modes we get
\begin{align}
  \hxi{\Bd \to \phi K^{*0}} & \leq 0.15 + 0.03\,\sqrt{10^6 \times
    \Br{\Bd \to \omega \omega}} \;,  \tag{\ref{eq:120}} 
\end{align}
\begin{align}
  \axi{\Bu \to \phi K^{*+}} & \leq 0.78\;,
  \tag{\ref{eq:123}} \\
  \axi{\Bu \to \rho^+ K^{*0}} & \leq 1.44\;, \tag{\ref{eq:127}} \\
  \axi{\Bu \to \rho^0 K^{*+}} & \leq 1.02\;, \tag{\ref{eq:128}}
\end{align}

All our constraints on the CP asymmetries in $b \to s$ transitions
agree quite well with the observed CP asymmetries. Some constraints
cannot be currently obtained. First, there are all the $b \to s$ modes
which have not been measured yet and are only bounded. Obviously, a
measurement of those is also needed if CP asymmetry is to be
measured. Second, all the $\Bd \to VP$ modes require a bound on some
$\Bd \to K^* K$ branching ratio. The most useful $K^*K$ mode is
different for each $\Bd \to VP$ mode. Third, the CP asymmetry of $\Bd
\to \phi K^{*0}$ (and $\Bd \to \omega K^{*0}$) require a bound on the
branching ratios of $\Bd \to \omega \omega$.

\begin{acknowledgments}
  I thank Yossi Nir and Guy Engelhard for helpful discussions and for
  their comments on the manuscript.
\end{acknowledgments}

\appendix

\section{Obtaining $SU(3)$ relations}
\label{sec:su3-formalism}

A simple way to obtain all $\su$ relations is to use tensor methods. In
this appendix we give the calculation details. The results agree with
the tables in~\cite{Grossman:2003qp} with the addition of $B_s$ modes
which we include here.

We write the singlet and octet pseudo-scalar and vector mesons as
$\su$ tensors of the appropriate rank
\begin{equation}
  \label{eq:129}
  P_1 = \eta_1 \;,
\end{equation}
\begin{eqnarray}
  \label{eq:130}
  (P_8)^i_j = 
  \begin{pmatrix}
    \frac{1}{\sqrt{2}}\, \pi^0 +\frac{1}{\sqrt{6}}\, \eta_8 & \pi^+ & K^+
    \\
    \pi^- & -\frac{1}{\sqrt{2}}\, \pi^0 + \frac{1}{\sqrt{6}}\, \eta_8 &
    K^0 \\
    K^- & \Kb & -\sqrt{\frac{2}{3}}\, \eta_8 
  \end{pmatrix} \;,
\end{eqnarray}
\begin{equation}
  \label{eq:131}
  V_1 = \phi_1\;,
\end{equation}
\begin{equation}
  \label{eq:132}
  (V_8)^i_j = 
  \begin{pmatrix}
    \frac{1}{\sqrt{2}}\, \rho^0 +\frac{1}{\sqrt{6}}\, \phi_8 & \rho^+ & {K^*}^+
    \\
    \rho^- & -\frac{1}{\sqrt{2}}\, \rho^0 + \frac{1}{\sqrt{6}}\, \phi_8 &
    {K^*}^0 \\
    {K^*}^- & \Ksb & -\sqrt{\frac{2}{3}}\, \phi_8 
  \end{pmatrix} \;,
\end{equation}
The $B$ mesons form a triplet of $\su$:
\begin{equation}
  \label{eq:133}
  (B_3)_i = 
  \begin{pmatrix}
    \Bu & \Bd & \Bs
  \end{pmatrix}\;.
\end{equation}

We combine the $b \to d$ and $b \to s$ Hamiltonian operators into
three rank $3$ tensors \cite{Savage:1989ub,Dighe:1995bm}:
\begin{align}\label{eq:134}
    ((H^q_3)^i)^j_k &= \left(
    \begin{pmatrix}
      0 & 0 & 0 \\
      0 & 0 & 0 \\
      0 & 0 & 0
    \end{pmatrix}\;,\right.
  \begin{pmatrix}
    \lambda^d_q & 0 & 0 \\
    0 & \lambda^d_q & 0 \\
    0 & 0 & \lambda^d_q
  \end{pmatrix}\;, \left.\begin{pmatrix}
      \lambda^s_q & 0 & 0 \\
      0 & \lambda^s_q & 0 \\
      0 & 0 & \lambda^s_q
    \end{pmatrix}
  \right)\;, \\
  \label{eq:135}
    ((H^q_{\overline{6}})^i)^j_k &= \left(
    \begin{pmatrix}
      0 & 0 & 0 \\
      \lambda^d_q & 0 & 0 \\
      \lambda^s_q & 0 & 0
    \end{pmatrix}\;,\right.
  \begin{pmatrix}
    -\lambda^d_q & 0 & 0 \\
    0 & 0 & 0 \\
    0 & -\lambda^s_q & \lambda^d_q
  \end{pmatrix}\;, \left.\begin{pmatrix}
      -\lambda^s_q & 0 & 0 \\
      0 & \lambda^s_q & -\lambda^d_q \\
      0 & 0 & 0
    \end{pmatrix}
  \right)\;, \\
  \label{eq:136}
    ((H^q_{15})^i)^j_k &= \left(
    \begin{pmatrix}
      0 & 0 & 0 \\
      3\lambda^d_q & 0 & 0 \\
      3\lambda^s_q & 0 & 0
    \end{pmatrix}\;,\right.
  \begin{pmatrix}
    3 \lambda^d_q & 0 & 0 \\
    0 & -2 \lambda^d_q & 0 \\
    0 & -\lambda^s_q & -\lambda^d_q
  \end{pmatrix}\;, \left. \begin{pmatrix}
      3 \lambda^s_q & 0 & 0 \\
      0 & - \lambda^s_q & -\lambda^d_q \\
      0 & 0 & -2 \lambda^s_q
    \end{pmatrix}
  \right)\;,
\end{align}
where $\lambda^{q^\prime}_q=V_{qb}^*V_{qq^\prime}$ and $q$ can be
either $u$ or $c$.

The effective Hamiltonian is obtained by contracting in all possible
ways the Hamiltonian operators with the meson tensors. We next proceed
and show how this is done for the relevant final states. 

\subsection{$B\to P_1 P_8$, $P_1 V_8$, $V_1 P_8$ and $V_1 V_8$}
\label{sec:bto-p_1-p_8-1}

In all the cases of $B\to P_1 P_8$, $P_1 V_8$, $V_1 P_8$ and $V_1
V_8$, the combination of meson representation is always an octet. The
effective Hamiltonian is therefore obtained by considering the following
contractions
\begin{equation}
  \label{eq:137}
  \begin{split}
    \mathcal{H}_\text{eff} &= A^{q8}_{15}\,(B_3)_i
    (H^q_{15})^{ij}_{k}\,(S_1)\,(M_8)^k_j \\
    & +
    A^{q8}_{~\overline{6}}\,(B_3)_i(H^q_{\overline{6}})^{ij}_{k}\,(S_1)\,(M_8)^k_j
    \\
    & + A^{q8}_{~3}\,(B_3)_i (H^q_3)^{ji}_{k}\,(S_1)\,(M_8)^k_j\;,
  \end{split}
\end{equation}
where both $S$ and $M$ can stand for either $P$ or $V$ and it is
understood that $q$ is summed over $q=u,\,c$. The
coefficients $A^{q8}_{~3}$, $A^{q8}_{~\overline{6}}$ and $A^{q8}_{15}$ are the reduced
matrix elements and can have, as far as the $\su$ analysis is
concerned, arbitrary values.

We point out, for clarity, that there are other ways to contract the
indices in~(\ref{eq:137}). For example we can change the $i$ and $j$
indices in $H_{\overline{6}}$ and $H_{15}$. However, one can easily
check that all other possible contractions give the same coefficients
as the three that already appear in~(\ref{eq:137}) and therefore
contribute to physical processes in the same way. There is, therefore,
no need to consider them.

\begingroup
\squeezetable
\begin{table}
  \centering
  \begin{equation*}
    \begin{array}{l||rrr}
       & {A^{q8}_{15}} & {A^{q8}_{~\overline{6}}} & {A^{q8}_{~3}} 
       \\ 
       \hline
       \hline
       {\Bd}\to\eta_1{K^0} &  -1 & -1 & 1 \\ 
       \hline
       {\Bu}\to\eta_1{K^+} & 3 & 1 & 1 \\ 
       \hline
       {\Bs}\to\eta_1{\pi^0} & 2\,{\sqrt{2}} & -{\sqrt{2}} & 0 \\ 
       {\Bs}\to\eta_1{\eta_8} & {\sqrt{6}} & 0 & -{\sqrt{\frac{2}{3}}}
       \\ 
       \hline
       \hline
       {\Bd}\to{\eta_1}{\pi^0} & \frac{5}{{\sqrt{2}}} & -
       \frac{1}{{\sqrt{2}}}  & - \frac{1}{{\sqrt{2}}}  \\   
       {\Bd}\to\eta_1{\eta_8} & {\sqrt{\frac{3}{2}}} &
       -{\sqrt{\frac{3}{2}}} & \frac{1}{{\sqrt{6}}} \\
       \hline
       {\Bu}\to\eta_1{\pi^+} & 3 & 1 & 1 \\
       \hline
       {\Bs}\to\eta_1{\Kb} & -1 &  -1 & 1 
\end{array}
  \end{equation*}
  \caption{$\su$ decomposition of $B\to S_1\,M_8$.}
  \label{tab:B-to-S_1M_8}
\end{table}
\endgroup

Expanding the effective Hamiltonian~(\ref{eq:137}) we arrange the
resulting numerical factors in table~\ref{tab:B-to-S_1M_8}. For
concreteness we list the $B\to P_1 P_8$ case, but the three other
cases can be obtained by simple substitution. We do not list the CKM
factors in table~\ref{tab:B-to-S_1M_8} since they can be easily
understood. For example, the $\Bu\to \eta_1 K^+$ decay amplitude is
obtained from the effective Hamiltonian term
\begin{equation}
  \label{eq:138}
  \Bu \eta_1 K^+ \left[\lambda^s_c\left(3\, A^{c8}_{15} +
      A^{c8}_{~\overline{6}}+A^{c8}_{~3}\right) + \lambda^s_u\left(3\, A^{u8}_{15} +
      A^{u8}_{~\overline{6}}+A^{u8}_{~3}\right)\right]\;.
\end{equation}

\subsection{$B\to P_8P_8$ and $B\to V_8V_8$.}
\label{sec:bto-p_8p_8-bto}

There are six ways (i.e. six representations) to combine two octets
into $\su$ invariant tensors:
\begin{align}
  (T^{MN}_1) &= (M_8)^m_n\, (N_8)^n_m \;, \label{eq:139} \\[1.5ex]
  (T^{MN}_{8s})^i_j &= \frac{1}{2}\left( (M_8)^i_m\, (N_8)^m_j + (N_8)^i_m\, (M_8)^m_j
  \right) -\frac{1}{3} (T^{MN}_1)\, \delta^i_j \;, \label{eq:140}
  \\[1.5ex]
  (T^{MN}_{8a})^i_j &= \frac{1}{2}\left( (M_8)^i_m\, (N_8)^m_j - (N_8)^i_m\, (M_8)^m_j
  \right) \;, \label{eq:141} \\[1.5ex]
  \begin{split}
    (T^{MN}_{10})_{ijk} &= \frac{1}{6} \left((M_8)^m_i (N_8)^n_j\,
      \varepsilon_{kmn} + (M_8)^m_j (N_8)^n_k\, \varepsilon_{imn} +
      (M_8)^m_k (N_8)^n_i \,\varepsilon_{jmn}\right. \\
    & \phantom{= \frac{1}{6}} + \left.(M_8)^m_k (N_8)^n_j\,
      \varepsilon_{imn} + (M_8)^m_j (N_8)^n_i\, \varepsilon_{kmn} +
      (M_8)^m_i (N_8)^n_k \,\varepsilon_{jmn} \right) \;, \label{eq:142}
  \end{split}
  \\[1.5ex]
  \begin{split}
    (T^{MN}_{\overline{10}})^{ijk} &= \frac{1}{6} \left((M_8)^i_m
      (N_8)^j_n \,\varepsilon^{kmn} + (M_8)^j_m (N_8)^k_n\,
      \varepsilon^{imn} + (M_8)^k_m (N_8)^i_n
      \,\varepsilon^{jmn}\right. \\
    & \phantom{= \frac{1}{6}}+\left.(M_8)^k_m
      (N_8)^j_n \,\varepsilon^{imn} + (M_8)^j_m (N_8)^i_n\,
      \varepsilon^{kmn} + (M_8)^i_m (N_8)^k_n
      \,\varepsilon^{jmn}\right) \;, \label{eq:143}
  \end{split}
  \\[1.5ex]
  \begin{split}
    (T^{MN}_{27})^{ij}_{kl} &= \frac{1}{4}\left( (M_8)^i_k (N_8)^j_l +
      (M_8)^j_k
      (N_8)^i_l + (M_8)^i_l (N_8)^j_k + (M_8)^j_l (N_8)^i_k\right)  \\
    & -\frac{1}{10}\left(\delta^i_k (T^{MN}_{8s})^j_l +
      \delta^i_l (T^{MN}_{8s})^j_k + \delta^j_k (T^{MN}_{8s})^i_l +
      \delta^j_l (T^{MN}_{8s})^i_k \right) \\
    &- \frac{1}{24}\left(
      \delta^i_k \delta^j_l + \delta^i_l \delta^j_k \right) (T^{MN}_1)
    \;.\label{eq:144}
  \end{split}
  \end{align}
Here both $M$ and $N$ stand for either $P$ or $V$.

When $M=N$, as in the $B\to P_8 P_8$ and $B\to V_8 V_8$ case, the only
non zero tensors are $(T^{MM}_{1})$, $(T^{MM}_{8s})$ and
$(T^{MM}_{27})$. Contacting them with the Hamiltonian operators and
the $B$ triplet we write
\begin{equation}
  \label{eq:145}
  \begin{split}
    \mathcal{H}_\text{eff} &= C^{q15}_{~27}\,(B_3)_i
    (H^q_{15})^{jk}_{l}\,(T^{MM}_{27})^{il}_{jk} +
    C^{q15}_{~8s}\,(B_3)_i
    (H^q_{15})^{ji}_{k}\,(T^{MM}_{8s})^{k}_{j} \\
    & + C^{q\overline{6}}_{8s}\,(B_3)_i
    (H^q_{\overline{6}})^{ji}_{k}\,(T^{MM}_{8s})^{k}_{j} \\
    & + C^{q3}_{8s}\,(B_3)_i (H^q_{3})^{ji}_{k}\,(T^{MM}_{8s})^{k}_{j}
    + C^{q3}_{~1}\,(B_3)_i (H^q_{3})^{ji}_{j}\,(T^{MM}_{1})\;.
  \end{split}
\end{equation}
Again, $M$ stands for either $P$ or $V$, and the index $q$ is summed
over $q=u,\,c$. As before, there are other ways to contract the
indices, which are equivalent to the ones we present and therefore
redundant.

The numerical factors are given in table~\ref{tab:B-to-M_8N_8}. We
write the mesons for the $B\to P_8P_8$ case while the $B\to V_8V_8$
case is obtained by simple substitution. One should note that if the
amplitudes are to be related to physical decay rates in a consistent
way, an additional factor of $\sqrt{2}$ needs to be introduced by hand
for final states which contain two identical mesons, such as
$\pi^0\pi^0$ or $\eta_8\eta_8$, due to the different phase space. Such
a factor was introduced in the table.

\begingroup
\squeezetable
\begin{table}
  \centering
  \begin{equation*}
    \begin{array}{l||rrrrr}
      & {C^{q15}_{~27}}
    & {C^{q15}_{~8s}} & {C^{q6}_{8s}
   } & {C^{q3}_{8s}} & {C^{q3}_{~1}
   } \\ 
   \hline
   \hline
   {\Bd \to}\,{K^0}\,
   {\eta_8} & \frac{4\,{\sqrt{6}}}
   {5} & \frac{1}{{\sqrt{6}}} & - 
      \frac{1}{{\sqrt{6}}}   & -
     \frac{1}{{\sqrt{6}}}   & 0 \\
   {\Bd \to}\,{K^0}\,
   {\pi^0} & \frac{12\,{\sqrt{2}}}
   {5} & \frac{1}{{\sqrt{2}}} & - 
      \frac{1}{{\sqrt{2}}}   & -
     \frac{1}{{\sqrt{2}}}   & 0 \\
     {\Bd \to}\,{K^+}\,
   {\pi^-} & \frac{16}{5} & 
    -1 & 1 & 1 & 0 \\ 
    \hline
    {\Bu \to}\,
   {K^+}\,{\pi^0} & 
    \frac{16\,{\sqrt{2}}}{5} & \frac{3}
   {{\sqrt{2}}} & - \frac{1}
     {{\sqrt{2}}}   & \frac{1}
   {{\sqrt{2}}} & 0 \\ 
   {\Bu \to}\,
   {K^+}\,{\eta_8} & 
    \frac{8\,{\sqrt{6}}}{5} & -{\sqrt{\frac{3}
       {2}}} & \frac{1}{{\sqrt{6}}} & -
     \frac{1}{{\sqrt{6}}}   & 0 \\
     {\Bu \to}\,{K^0}\,
   {\pi^+} & - \frac{8}
     {5}   & 3 & -1 & 1 & 0 \\ 
     \hline
   {\Bs \to}\,{K^0}\,
   {\Kb} & - \frac{2}{5}
       & -3 & -1 & \frac{1}
   {3} & 2 \\ {\Bs \to}\,
   {K^-}\,{K^+} & \frac{14}
   {5} & 1 & 1 & \frac{1}{3} & 2 \\ 
   {\Bs \to}\,{{\pi^0}{\pi^0}} &
   - \frac{\sqrt{2}}{5} 
      & \sqrt{2} & 0 & - \frac{\sqrt{2}}{3} 
      & \sqrt{2} \\ {\Bs \to}\,
   {\eta_8}\,{\pi^0} &
   \frac{-8\,{\sqrt{3}}}{5} & \frac{4}
   {{\sqrt{3}}} & \frac{2}
   {{\sqrt{3}}} & 0 & 0 \\ {\Bs \to}\,
   {{\eta_8}{\eta_8}} & - \frac{9\sqrt{2}}
     {5}   & -\sqrt{2} & 0 & \frac{\sqrt{2}}
   {3} & \sqrt{2} \\ {\Bs \to}\,
   {\pi^-}\,{\pi^+} &
   - \frac{2}{5} 
      & 2 & 0 & - \frac{2}{3} 
      & 2 \\ 
      \hline
      \hline
      {\Bd \to}\,{K^0}\,
   {\Kb} & - \frac{2}{5}
       & -3 & -1 & \frac{1}
   {3} & 2 \\ {\Bd \to}\,
   {K^-}\,{K^+} & -
     \frac{2}{5}   & 2 & 0 & -
     \frac{2}{3}   & 2 \\ {\Bd \to}\,{{\pi^0}{\pi^0}} & - 
      \frac{13\sqrt{2}}{5}   & \frac{\sqrt{2}}
   {2} & \frac{\sqrt{2}}{2} & \frac{\sqrt{2}}
   {6} & \sqrt{2} \\ {\Bd \to}\,
   {\eta_8}\,{\pi^0} & 
   0 & \frac{5}{{\sqrt{3}}} & \frac{1}
   {{\sqrt{3}}} & - \frac{1}
     {{\sqrt{3}}}   & 0 \\ {
    \Bd \to}\,{{\eta_8}{\eta_8}} & \frac{3\sqrt{2}}
   {5} & - \frac{1}{\sqrt{2}}   & - \frac{1}{\sqrt{2}}   & - \frac{\sqrt{2}}
     {6}   & \sqrt{2} \\ {\Bd \to}\,
   {\pi^-}\,{\pi^+} &
   \frac{14}{5} & 1 & 1 & \frac{1}
   {3} & 2 \\ 
   \hline
   {\Bu \to}\,
   {\Kb}\,{K^+} & -
     \frac{8}{5}   & 3 & 
    -1 & 1 & 0 \\ {\Bu \to}\,
   {\pi^0}\,{\pi^+} &
   4\,{\sqrt{2}} & 0 & 0 & 0 & 0 \\ 
   {\Bu \to}\,{\eta_8}\,
   {\pi^+} & \frac{4\,{\sqrt{6}}}
   {5} & {\sqrt{6}} & -{\sqrt{\frac{2}
       {3}}} & {\sqrt{\frac{2}
      {3}}} & 0 \\ 
  \hline
  {\Bs \to}\,
   {\Kb}\,{\pi^0} & 
    \frac{12\,{\sqrt{2}}}{5} & \frac{1}
   {{\sqrt{2}}} & - \frac{1}
     {{\sqrt{2}}}   & - \frac{1}
     {{\sqrt{2}}}   & 0 \\ {
    \Bs \to}\,{\Kb}\,{\eta_8} &
   \frac{4\,{\sqrt{6}}}{5} & \frac{1}
   {{\sqrt{6}}} & - \frac{1}
     {{\sqrt{6}}}   & - \frac{1}
     {{\sqrt{6}}}   & 0 \\ {
    \Bs \to}\,{K^-}\,{\pi^+} &
   \frac{16}{5} & -1 & 1 & 1 & 0 
    \end{array}
  \end{equation*}
  \caption{$\su$ decomposition of $B\to M_8N_8$.}
  \label{tab:B-to-M_8N_8}
\end{table}
\endgroup

\subsection{$B\to P_8V_8$.}
\label{sec:bto-p_8v_8}

Setting $M=P$ and $N=V$, all six invariant
tensors~(\ref{eq:139})-(\ref{eq:144}) are now non zero. We write the fully
contracted Hamiltonian by adding the necessary terms to~(\ref{eq:145})
\begin{equation}
  \label{eq:146}
  \begin{split}
    \mathcal{H}_\text{eff} &= E^{q15}_{~27}\,(B_3)_i
    (H^q_{15})^{jk}_{l}\,(T^{PV}_{27})^{il}_{jk} +
    E^{q15}_{~8s}\,(B_3)_i
    (H^q_{15})^{ji}_{k}\,(T^{PV}_{8s})^{k}_{j} \\
    & + E^{q\overline{6}}_{8s}\,(B_3)_i
    (H^q_{\overline{6}})^{ji}_{k}\,(T^{PV}_{8s})^{k}_{j} \\
    & + E^{q3}_{8s}\,(B_3)_i (H^q_{3})^{ji}_{k}\,(T^{PV}_{8s})^{k}_{j}
    + E^{q3}_{~1}\,(B_3)_i (H^q_{3})^{ji}_{j}\,(T^{PV}_{1}) \\
    & + E^{q15}_{10}\,(B_3)_i
    (H^q_{15})^{jk}_{l}\,(T^{PV}_{10})_{jkm}(\varepsilon)^{mli} +
    E^{q15}_{~8a}\,(B_3)_i
    (H^q_{15})^{ji}_{k}\,(T^{PV}_{8a})^{k}_{j} \\
    & + E^{q\overline{6}}_{\overline{10}}\,(B_3)_i
    (H^q_{\overline{6}})^{jk}_{l}\,(T^{PV}_{\overline{10}})^{ilm}(\varepsilon)_{jkm}
    + E^{q\overline{6}}_{8a}\,(B_3)_i (H^q_{\overline{6}})^{ji}_{k}\,(T^{PV}_{8s})^{k}_{j}
    \\
    & + E^{q3}_{8a}\,(B_3)_i (H^q_{3})^{ji}_{K}\,(T^{PV}_{8a})^k_j \;.
  \end{split}
\end{equation}
Again $q$ is summed over $q=u,\,c$ and we do not write other possible
contractions which contribute in the same way. The numerical factors
are given in table~\ref{tab:B-to-P_8V_8}.

\begingroup
\squeezetable
\begin{table}
  \centering
  \begin{equation*}
    \begin{array}{l||rrrrrrrrrr}
      & {E^{q15}_{~27}
      } & {E^{q15}_{~8s}} & {E^{q\overline{6}}_{8s}
      } & {E^{q3}_{8s}} & {E^{q3}_{~1}
      } & {E^{q15}_{~10}} & {E^{q\overline{6}}_{\overline{10}}
      } & {E^{q15}_{~8a}} & {E^{q6}_{8a}
      } & {E^{q3}_{8a}} \\ 
      \hline
      \hline
      {\Bd \to}\,
      {K^0}\,{\phi_8} & 
      \frac{2\,{\sqrt{6}}}{5} & \frac{1}
      {{\sqrt{6}}} & \frac{-1}
      {2\,{\sqrt{6}}} & \frac{-1}
      {2\,{\sqrt{6}}} & 0 & -4\,
      {\sqrt{\frac{2}{3}}} & 0 & \frac{{
          \sqrt{\frac{3}{2}}}}{2} & \frac{-{
          \sqrt{\frac{3}{2}}}}{2} & \frac{-
        {\sqrt{\frac{3}{2}}}}{2} \\ 
      {\Bd \to}\,{K^0}\,{\rho^0} &
      \frac{6\,{\sqrt{2}}}{5} & \frac{1}
      {{\sqrt{2}}} & \frac{-1}
      {2\,{\sqrt{2}}} & \frac{-1}
      {2\,{\sqrt{2}}} & 0 & \frac{-4\,{\sqrt{2}}}
      {3} & \frac{-4\,{\sqrt{2}}}{3} & \frac{-1}
      {2\,{\sqrt{2}}} & \frac{1}
      {2\,{\sqrt{2}}} & \frac{1}
      {2\,{\sqrt{2}}} \\ {\Bd \to}\,
      {\eta_8}\,{K^{*0}} & 
      \frac{2\,{\sqrt{6}}}{5} & \frac{1}
      {{\sqrt{6}}} & \frac{-1}
      {2\,{\sqrt{6}}} & \frac{-1}
      {2\,{\sqrt{6}}} & 0 & 4\,
      {\sqrt{\frac{2}{3}}} & 0 & \frac{-{
          \sqrt{\frac{3}{2}}}}{2} & \frac{{
          \sqrt{\frac{3}{2}}}}{2} & \frac{{
          \sqrt{\frac{3}{2}}}}{2} \\ {
        \Bd \to}\,{\pi^0}\,{K^{*0}} &
      \frac{6\,{\sqrt{2}}}{5} & \frac{1}
      {{\sqrt{2}}} & \frac{-1}
      {2\,{\sqrt{2}}} & \frac{-1}
      {2\,{\sqrt{2}}} & 0 & \frac{4\,{\sqrt{2}}}
      {3} & \frac{4\,{\sqrt{2}}}{3} & \frac{1}
      {2\,{\sqrt{2}}} & \frac{-1}
      {2\,{\sqrt{2}}} & \frac{-1}
      {2\,{\sqrt{2}}} \\ {\Bd \to}\,{\pi^-}\,
      {K^{*+}} & 
      \frac{8}{5} & -1 & \frac{1}{2} & \frac{1}
      {2} & 0 & - \frac{8}{3} 
      & \frac{4}{3} & - \frac{1}{2}
      & \frac{1}{2} & \frac{1}
      {2} \\ {\Bd \to}\,
      {\rho^-}\,{K^+} & \frac{8}{5} & 
      -1 & \frac{1}{2} & \frac{1}{2} & 0 & 
      \frac{8}{3} & - \frac{4}{3} 
      & \frac{1}{2} & - \frac{1}{2}
      & - \frac{1}{2} 
      \\ 
      \hline
      {\Bu \to}\,
      {\pi^0}\,{K^{*+}} & \frac{8\,{\sqrt{2}}}
      {5} & 0 & \frac{-1}{2\,{\sqrt{2}}} & 
      \frac{1}{2\,{\sqrt{2}}} & 0 & 0 & 
      \frac{4\,{\sqrt{2}}}{3} & \frac{3}
      {2\,{\sqrt{2}}} & \frac{-1}
      {2\,{\sqrt{2}}} & \frac{1}
      {2\,{\sqrt{2}}} \\ {\Bu \to}\,{\eta_8}\,
      {K^{*+}} & 
      \frac{4\,{\sqrt{6}}}{5} & 0 & \frac{1}
      {2\,{\sqrt{6}}} & \frac{-1}
      {2\,{\sqrt{6}}} & 0 & 0 & 0 & \frac{3\,
        {\sqrt{\frac{3}{2}}}}{2} & \frac{-{
          \sqrt{\frac{3}{2}}}}{2} & \frac{{
          \sqrt{\frac{3}{2}}}}{2} \\ {
        \Bu \to}\,{\pi^+}\,{K^{*0}} &
      - \frac{4}{5}   & 0 & -
      \frac{1}{2}   & \frac{1}
      {2} & 0 & 0 & - \frac{4}{3} 
      & \frac{3}{2} & - \frac{1}{2}
      & \frac{1}{2} \\ {\Bu \to
      }\,{K^+}\,{\rho^0} &
      \frac{8\,{\sqrt{2}}}{5} & 0 & \frac{-1}
      {2\,{\sqrt{2}}} & \frac{1}
      {2\,{\sqrt{2}}} & 0 & 0 & \frac{-4\,
        {\sqrt{2}}}{3} & \frac{-3}
      {2\,{\sqrt{2}}} & \frac{1}
      {2\,{\sqrt{2}}} & \frac{-1}
      {2\,{\sqrt{2}}} \\ {\Bu \to}\,
      {K^+}\,{\phi_8} & 
      \frac{4\,{\sqrt{6}}}{5} & 0 & \frac{1}
      {2\,{\sqrt{6}}} & \frac{-1}
      {2\,{\sqrt{6}}} & 0 & 0 & 0 & \frac{-3\,
        {\sqrt{\frac{3}{2}}}}{2} & \frac{{
          \sqrt{\frac{3}{2}}}}{2} & \frac{-{
          \sqrt{\frac{3}{2}}}}{2} \\ 
      {\Bu \to}\,{K^0}\,{\rho^+} &
      - \frac{4}{5}   & 0 & -
      \frac{1}{2}   & \frac{1}
      {2} & 0 & 0 & \frac{4}{3} & - 
      \frac{3}{2}   & \frac{1}
      {2} & - \frac{1}{2} 
      \\ 
      \hline
      {\Bs \to}\,{\Kb}\,
      {K^{*0}} & - \frac{1}{5}
      & - \frac{1}{3} 
      & - \frac{1}{2}   & 
      \frac{1}{6} & 1 & \frac{4}{3} & -
      \frac{2}{3}   & - \frac{1}
      {2}   & \frac{1}{2} & \frac{1}
      {2} \\ {\Bs \to}\,{K^-}\,
      {K^{*+}} & \frac{7}{5} & -
      \frac{1}{3}   & \frac{1}
      {2} & \frac{1}{6} & 1 & - \frac{4}
      {3}   & \frac{2}{3} & -
      \frac{5}{2}   & - \frac{1}
      {2}   & \frac{1}{2} \\ 
      {\Bs \to}\,{\eta_8}\,
      {\rho^0} & \frac{-4\,{\sqrt{3}}}
      {5} & 0 & \frac{1}
      {{\sqrt{3}}} & 0 & 0 & \frac{4}
      {{\sqrt{3}}} & \frac{2}
      {{\sqrt{3}}} & 0 & 0 & 0 \\ {\Bs \to
      }\,{\pi^-}\,{\rho^+} &
      - \frac{1}{5}   & \frac{2}
      {3} & 0 & - \frac{1}{3} 
      & 1 & \frac{4}{3} & - \frac{2}
      {3}   & -2 & -1 & 0 \\ 
      {\Bs \to}\,{K^0}\,{\Ksb} &
      - \frac{1}{5}   & -
      \frac{1}{3}   & - \frac{1}
      {2}   & \frac{1}{6} & 1 & -
      \frac{4}{3}   & \frac{2}
      {3} & \frac{1}{2} & - \frac{1}
      {2}   & - \frac{1}{2}
      \\ {\Bs \to}\,
      {K^+}\,{K^{*-}} & \frac{7}
      {5} & - \frac{1}{3}   & 
      \frac{1}{2} & \frac{1}{6} & 1 & \frac{4}
      {3} & - \frac{2}{3}   & 
      \frac{5}{2} & \frac{1}{2} & - 
      \frac{1}{2}   \\ {\Bs \to
      }\,{\pi^0}\,{\rho^0} &
      - \frac{1}{5}   & \frac{2}
      {3} & 0 & - \frac{1}{3} 
      & 1 & 0 & 0 & 0 & 0 & 0 \\ {\Bs \to}\,{\pi^0}\,{\phi_8}
      & \frac{-4\,{\sqrt{3}}}{5} & 0 & \frac{1}
      {{\sqrt{3}}} & 0 & 0 & \frac{-4}
      {{\sqrt{3}}} & \frac{-2}
      {{\sqrt{3}}} & 0 & 0 & 0 \\ {\Bs \to
      }\,{\eta_8}\,{\phi_8}
      & - \frac{9}{5}   & -
      \frac{2}{3}   & 0 & \frac{1}
      {3} & 1 & 0 & 0 & 0 & 0 & 0 \\ {
        \Bs \to}\,{\pi^+}\,{\rho^-}
      & - \frac{1}{5}   & \frac{2}
      {3} & 0 & - \frac{1}{3} 
      & 1 & - \frac{4}{3} 
      & \frac{2}{3} & 2 & 1 & 0 \\ 
      \hline
      \hline
      {\Bd \to}\,{K^0}\,{\Ksb} &
      - \frac{1}{5}   & -
      \frac{1}{3}   & - \frac{1}
      {2}   & \frac{1}{6} & 1 & \frac{4}
      {3} & - \frac{2}{3}   & -\frac{1}{2}   & \frac{1}
      {2} & \frac{1}{2} \\ {\Bd \to}\,
      {K^+}\,{K^{*-}} & -
      \frac{1}{5}   & \frac{2}
      {3} & 0 & - \frac{1}{3} 
      & 1 & - \frac{4}{3} 
      & \frac{2}{3} & 2 & 1 & 0 \\ 
      {\Bd \to}\,{\pi^0}\,
      {\rho^0} & - \frac{13}
      {5}   & - \frac{1}{3}
      & \frac{1}{2} & \frac{1}
      {6} & 1 & 0 & 0 & 0 & 0 & 0 \\ {
        \Bd \to}\,{\pi^0}\,{\phi_8}
      & 0 & \frac{1}{{\sqrt{3}}} & \frac{1}
      {2\,{\sqrt{3}}} & \frac{-1}
      {2\,{\sqrt{3}}} & 0 & \frac{4}
      {{\sqrt{3}}} & \frac{2}
      {{\sqrt{3}}} & 0 & 0 & 0 \\ {\Bd \to
      }\,{\eta_8}\,{\phi_8}
      & \frac{3}{5} & \frac{1}{3} & - 
      \frac{1}{2}   & - \frac{1}
      {6}   & 1 & 0 & 0 & 0 & 0 & 0 \\
      {\Bd \to}\,{\pi^+}\,
      {\rho^-} & \frac{7}{5} & -
      \frac{1}{3}   & \frac{1}
      {2} & \frac{1}{6} & 1 & \frac{4}{3} & 
      - \frac{2}{3}   & \frac{5}
      {2} & \frac{1}{2} & - \frac{1}
      {2}   \\ {\Bd \to}\,
      {\Kb}\,{K^{*0}} & -
      \frac{1}{5}   & - \frac{1}
      {3}   & - \frac{1}{2}
      & \frac{1}{6} & 1 & -
      \frac{4}{3}   & \frac{2}
      {3} & \frac{1}{2} & - \frac{1}
      {2}   & - \frac{1}{2}
      \\ {\Bd \to}\,
      {K^-}\,{K^{*+}} & -
      \frac{1}{5}   & \frac{2}
      {3} & 0 & - \frac{1}{3} 
      & 1 & \frac{4}{3} & - \frac{2}
      {3}   & -2 & -1 & 0 \\ 
      {\Bd \to}\,{\eta_8}\,
      {\rho^0} & 0 & \frac{1}
      {{\sqrt{3}}} & \frac{1}
      {2\,{\sqrt{3}}} & \frac{-1}
      {2\,{\sqrt{3}}} & 0 & \frac{-4}
      {{\sqrt{3}}} & \frac{-2}
      {{\sqrt{3}}} & 0 & 0 & 0 \\ {\Bd \to
      }\,{\pi^-}\,{\rho^+} &
      \frac{7}{5} & - \frac{1}{3} 
      & \frac{1}{2} & \frac{1}{6} & 1 & -\frac{4}{3}   & \frac{2}
      {3} & - \frac{5}{2}   & -\frac{1}{2}   & \frac{1}
      {2} \\ 
      \hline
      {\Bu \to}\,{K^+}\,
      {\Ksb} & - \frac{4}{5}
      & 0 & - \frac{1}{2}
      & \frac{1}{2} & 0 & 0 & -
      \frac{4}{3}   & \frac{3}
      {2} & - \frac{1}{2}   & 
      \frac{1}{2} \\ {\Bu \to}\,
      {\pi^0}\,{\rho^+} &
      2\,{\sqrt{2}} & 0 & 0 & 0 & 0 & 0 & 
      \frac{2\,{\sqrt{2}}}{3} & \frac{3}
      {{\sqrt{2}}} & - \frac{1}
      {{\sqrt{2}}}   & \frac{1}
      {{\sqrt{2}}} \\ {\Bu \to}\,
      {\eta_8}\,{\rho^+} &
      \frac{2\,{\sqrt{6}}}{5} & 0 & - 
      \frac{1}{{\sqrt{6}}}   & \frac{1}
      {{\sqrt{6}}} & 0 & 0 & 2\,
      {\sqrt{\frac{2}{3}}} & 0 & 0 & 0 \\ 
      {\Bu \to}\,{\Kb}\,
      {K^{*+}} & - \frac{4}{5}
      & 0 & - \frac{1}{2}
      & \frac{1}{2} & 0 & 0 & \frac{4}
      {3} & - \frac{3}{2}   & 
      \frac{1}{2} & - \frac{1}{2} 
      \\ {\Bu \to}\,{\pi^+}\,
      {\rho^0} & 2\,
      {\sqrt{2}} & 0 & 0 & 0 & 0 & 0 & \frac
      {-2\,{\sqrt{2}}}{3} & \frac{-3}
      {{\sqrt{2}}} & \frac{1}{{\sqrt{2}}} & 
      - \frac{1}{{\sqrt{2}}}   \\
      {\Bu \to}\,{\pi^+}\,
      {\phi_8} & \frac{2\,{\sqrt{6}}}
      {5} & 0 & - \frac{1}{{\sqrt{6}}}
      & \frac{1}
      {{\sqrt{6}}} & 0 & 0 & -2\,
      {\sqrt{\frac{2}{3}}} & 0 & 0 & 0 \\ 
      \hline
      {\Bs \to}\,{\Kb}\,
      {\rho^0} & \frac{6\,{\sqrt{2}}}
      {5} & \frac{1}{{\sqrt{2}}} & \frac{-1}
      {2\,{\sqrt{2}}} & \frac{-1}
      {2\,{\sqrt{2}}} & 0 & \frac{-8\,{\sqrt{2}}}
      {3} & \frac{-2\,{\sqrt{2}}}{3} & \frac{1}
      {2\,{\sqrt{2}}} & \frac{-1}
      {2\,{\sqrt{2}}} & \frac{-1}
      {2\,{\sqrt{2}}} \\ {\Bs \to}\,
      {\Kb}\,{\phi_8} & 
      \frac{2\,{\sqrt{6}}}{5} & \frac{1}
      {{\sqrt{6}}} & \frac{-1}
      {2\,{\sqrt{6}}} & \frac{-1}
      {2\,{\sqrt{6}}} & 0 & 0 & -2\,
      {\sqrt{\frac{2}{3}}} & \frac{-{\sqrt{
            \frac{3}{2}}}}{2} & \frac{{
          \sqrt{\frac{3}{2}}}}{2} & \frac{{
          \sqrt{\frac{3}{2}}}}{2} \\ {
        \Bs \to}\,{K^-}\,{\rho^+} &
      \frac{8}{5} & -1 & \frac{1}{2} & \frac{1}
      {2} & 0 & - \frac{8}{3} 
      & \frac{4}{3} & - \frac{1}{2}
      & \frac{1}{2} & \frac{1}
      {2} \\ {\Bs \to}\,
      {\pi^0}\,{\Ksb} & \frac{6\,{\sqrt{2}}}
      {5} & \frac{1}{{\sqrt{2}}} & \frac{-1}
      {2\,{\sqrt{2}}} & \frac{-1}
      {2\,{\sqrt{2}}} & 0 & \frac{8\,{\sqrt{2}}}
      {3} & \frac{2\,{\sqrt{2}}}{3} & \frac{-1}
      {2\,{\sqrt{2}}} & \frac{1}
      {2\,{\sqrt{2}}} & \frac{1}
      {2\,{\sqrt{2}}} \\ {\Bs \to}\,{\eta_8}\,
      {\Ksb} &
      \frac{2\,{\sqrt{6}}}{5} & \frac{1}
      {{\sqrt{6}}} & \frac{-1}
      {2\,{\sqrt{6}}} & \frac{-1}
      {2\,{\sqrt{6}}} & 0 & 0 & 2\,
      {\sqrt{\frac{2}{3}}} & \frac{{\sqrt{
            \frac{3}{2}}}}{2} & \frac{-{
          \sqrt{\frac{3}{2}}}}{2} & \frac{-
        {\sqrt{\frac{3}{2}}}}{2} \\ 
      {\Bs \to}\,{\pi^+}\,{K^{*-}} &
      \frac{8}{5} & -1 & \frac{1}{2} & \frac{1}
      {2} & 0 & \frac{8}{3} & - \frac{4}
      {3}   & \frac{1}{2} & -
      \frac{1}{2}   & - \frac{1}
      {2}   
    \end{array}
  \end{equation*}
  \caption{$\su$ decomposition of $B\to P_8V_8$.}
  \label{tab:B-to-P_8V_8}
\end{table}
\endgroup

\section{Experimental data}
\label{sec:experimental-data}

We collect in this appendix the current experimental
data~\cite{HFAG}. All branching ratios are given in units of
$10^{-6}$. Since we only list $\Bu$ and $\Bd$ branching ratios,
the identity of the decaying meson is self evident.

\subsection{$B \to PP$, $b \to s$ modes}
\label{sec:b-pp-delta}

\begin{align}
  \label{eq:147}
  \Br{\eta K^0} & < 1.9\;, \\
  \Br{\eta' K^0} & = 63.2\pm 3.3\;, \\
  \Br{\pi^0 K^0} & = 11.5\pm 1.0 \;, \\
  \Br{\pi^- K^+} &= 18.9\pm 0.7 \;,  
\end{align}
\begin{align}
\Br{\eta K^+} & = 2.5 \pm 0.3 \;, \\
  \Br{\eta' K^+} & = 69.4\pm 2.7 \;, \\
  \Br{\pi^0 K^+} & = 12.1 \pm 0.8 \;, \\
  \Br{\pi^+ K^0} & = 24.1\pm 1.3 \;.
\end{align}

\subsection{$B \to PP$, $b \to d$ modes}
\label{sec:b-to-pp-1}

\begin{align}
  \Br{\eta \eta} & < 2.0 \;, \\
  \Br{\eta' \eta'} & < 10 \;, \\
  \Br{\eta' \eta} & < 4.6 \;, \\
  \Br{\pi^0 \pi^0} & = 1.45\pm 0.29 \;, \\
  \Br{\pi^+ \pi^-} & = 5.0\pm 0.4 \;, \\
  \Br{\eta \pi^0} & < 2.5 \;, \\
  \Br{\eta' \pi^0} & < 3.7 \;, \\
  \Br{K^0 \Kb} & = 0.96^{+0.25}_{-0.24} \;, \\
  \Br{K^+ K^-} & = 0.05^{+0.10}_{-0.09} \;, 
\end{align}
\begin{align}
  \Br{\eta \pi^+} & = 4.3\pm 0.4 \;, \\
  \Br{\eta' \pi^+} & = 2.53^{+0.59}_{-0.50} \;, \\
  \Br{\pi^0 \pi^+} & = 5.5\pm 0.6 \;, \\
  \Br{\Kb K^+} & = 1.2\pm 0.3 \;.
\end{align}

\subsection{$B \to V P$, $b \to s$ modes}
\label{sec:b-to-pv}

\begin{align}
  \Br{\phi K^0} & = 8.3^{+1.2}_{-1.0} \;, \\
  \Br{\omega K^0} & = 4.7\pm 0.6 \;, \\
  \Br{\rho^0 K^0} & = 5.1\pm 1.6 \;, \\
  \Br{\rho^- K^+} & = 9.9^{+1.6}_{-1.5} \;, \\
  \Br{K^{*0} \eta} & = 18.7\pm 1.7 \;, \\
  \Br{K^{*0} \eta'} & < 7.6 \;, \\
  \Br{K^{*0} \pi^0} & = 1.7\pm 0.8 \;, \\
  \Br{K^{*+} \pi^-} & = 11.7^{+1.5}_{-1.4} \;,
\end{align}
\begin{align}
  \Br{\phi K^+} & = 9.03^{+0.65}_{-0.63} \;, \\
  \Br{\omega K^+} & = 6.5 \pm 0.6 \;, \\
  \Br{\rho^0 K^+} & = 4.23^{+0.56}_{-0.57}\;, \\
  \Br{\rho^+ K^0} & < 48 \;, \\
  \Br{K^{*+} \eta} & = 24.3^{+3.0}_{2.9} \;, \\
  \Br{K^{*+} \eta'} & < 14 \;, \\
  \Br{K^{*+} \pi^0} &= 6.9 \pm 2.3 \;, \\
  \Br{K^{*0} \pi^+} & = 10.8\pm 0.8 \;.
\end{align}

\subsection{$B \to V P$, $b \to d$ modes}
\label{sec:b-to-v}

\begin{align}
  \Br{\phi \eta} & < 1.0 \;, \\
  \Br{\phi \eta'} &  < 4.5 \;, \\
  \Br{\phi \pi^0} &  < 1.0 \;, \\
  \Br{\omega \eta} & < 1.9 \;, \\
  \Br{\omega \eta'} & < 2.8 \;, \\
  \Br{\omega \pi^0} &  < 1.2 \;, \\
  \Br{\rho^0 \eta} &  < 1.5 \;, \\
  \Br{\rho^0 \eta'} & < 4.3 \;, \\
  \Br{\rho^0 \pi^0} &  = 1.83^{+0.56}_{-0.55} \;, \\
  \Br{\rho^+ \pi^-} &  = 24.0\pm 2.5 \;, \\
  \Br{\rho^- \pi^+} &  = 24.0\pm 2.5 \;, \\
  \Br{K^{*0} \Kb} & < \text{Not measured yet} \;, \\
  \Br{\Ksb K^0} & < \text{Not measured yet} \;, \\
  \Br{K^{*+} K^-} & < \text{Not measured yet} \;, \\
  \Br{K^{*-} K^+} & < \text{Not measured yet} \;, 
\end{align}
\begin{align}
  \Br{\phi \pi^+} & < 0.41 \;, \\
  \Br{\omega \pi^+} & = 6.6\pm 0.6 \;, \\
  \Br{\rho^0 \pi^+} & = 8.7^{+1.0}_{-1.1} \;, \\
  \Br{\rho^+ \eta} & = 8.1^{+1.7}_{-1.5} \;, \\
  \Br{\rho^+ \eta'} & < 22 \;, \\
  \Br{\rho^+ \pi^0} & = 10.8^{+1.4}_{-1.5} \;,  \\
  \Br{\Ksb K^+} & < 5.3 \;, \\
  \Br{K^{*+} \Kb} & = \text{Not measured yet} \;.
\end{align}

\subsection{$B \to VV$, $b \to s$ modes}
\label{sec:b-to-vv}

\begin{align}
  \Br{\phi K^{*0}} & = 9.5 \pm 0.9 \;, \\
  \Br{\omega K^{*0}} & < 6.0 \;, \\
  \Br{\rho^0 K^{*0}} & < 2.6 \;, \\
  \Br{\rho^- K^{*+}} & < 24 \;,  
\end{align}
\begin{align}
\Br{\phi K^{*+}} & = 9.7 \pm 1.5 \;, \\
  \Br{\omega K^{*+}} & < 7.4 \;, \\
  \Br{\rho^0 K^{*+}} & = 10.6^{+3.8}_{-3.5} \;, \\
  \Br{\rho^+ K^{*0}} & = 10.6 \pm 1.9 \;.
\end{align}

\subsection{$B \to VV$, $b \to d$ modes}
\label{sec:b-to-vv-1}

\begin{align}
  \Br{\phi \phi} & < 1.5 \;, \\
  \Br{\omega \omega} & < \text{Not measured yet} \;, \\
  \Br{\phi \omega} & < \text{Not measured yet} \;, \\
  \Br{\rho^0 \rho^0} & < 1.1 \;, \\
  \Br{\rho^+ \rho^-} & = 26.2^{+3.6}_{-3.7} \;, \\
  \Br{\phi \rho^0} & < 13 \;, \\
  \Br{\omega \rho^0} & < 3.3 \;, \\
  \Br{K^{*0} \Ksb} & < 22 \;, \\
  \Br{K^{*+} K^{*-}} & < 141 \;, 
\end{align}
\begin{align}
  \Br{\phi \rho^+} & < 16 \;, \\
  \Br{\omega \rho^+} & = 12.6^{+4.0}_{-3.7} \;, \\
  \Br{\rho^0 \rho^+} & = 26.4^{+6.1}_{-6.4} \;, \\
  \Br{\Ksb K^{*+}} & < 71 \;.
\end{align}

\end{document}